\documentclass[aps,pra,10pt,twocolumn,amsfonts,superscriptaddress,showpacs,floatfix]{revtex4-1}
\pdfoutput=1
\usepackage{comment, bbm}
\usepackage{enumerate}
\usepackage{amssymb}
\usepackage{amsmath}
\usepackage{braket}
\usepackage{graphicx}
\usepackage{mathtools}
\usepackage{float}
\usepackage[usenames,dvipsnames]{color}
\usepackage[colorlinks,bookmarks=false,citecolor=NavyBlue,linkcolor=OliveGreen,url
color=blue]{hyperref}
\usepackage{tensor}
\usepackage{standalone}
\usepackage{scalerel}
\usepackage{tikz}
\usetikzlibrary{tikzmark,calc}
\usetikzlibrary{arrows.meta}
\usepackage{multirow}
\usetikzlibrary{positioning}
\usetikzlibrary{decorations.pathreplacing,angles,quotes, calligraphy}
\newcommand{\be}{\begin{equation}}
\newcommand{\ee}{\end{equation}}
\newcommand{\ba}{\begin{aligned}}
\newcommand{\ea}{\end{aligned}}
\newcommand{\bw}{\begin{widetext}}
\newcommand{\ew}{\end{widetext}}
\newcommand{\1}{\mathbbm{1}}

\begin{document}
\title{Statistics of the Spectral Form Factor in the Self-Dual Kicked Ising Model}
\author{Ana Flack}
\author{Bruno Bertini}
\author{Toma\v z Prosen}
\affiliation{Department of Physics, Faculty of Mathematics and Physics, University of Ljubljana, Jadranska 19, SI-1000 Ljubljana, Slovenia}
\begin{abstract}
We compute the full probability distribution of the spectral form factor in the self-dual kicked Ising model by providing an exact lower bound for each moment and verifying numerically that the latter is saturated. We show that at large enough times the probability distribution agrees exactly with the prediction of Random Matrix Theory if one identifies the appropriate ensemble of random matrices. We find that this ensemble is not the circular orthogonal one --- composed of symmetric random unitary matrices and associated with time-reversal-invariant evolution operators --- but is an ensemble of random matrices on a more restricted symmetric space (depending on the parity of the number of sites this space is either ${Sp(N)/U(N)}$ or ${O(2N)/{O(N)\!\times\!O(N)}}$). Even if the latter ensembles yield the same averaged spectral form factor as the circular orthogonal ensemble they show substantially enhanced fluctuations. This behaviour is due to a recently identified additional 
anti-unitary symmetry of the self-dual kicked Ising model.      
\end{abstract}
\maketitle

\section{Introduction}

The \emph{quantum chaos conjecture}~\cite{CGV80,Berry81,BGS84} states that a quantum system is \emph{chaotic} if the correlations of its energy levels have the same structure as those of random hermitian matrices~\cite{Haake, Mehtabook}. This ``conjecture" originates from studies on single-particle quantum systems, where the aforementioned property can be connected to the conventional chaoticity of the system (i.e. sensitivity of system's trajectories to initial conditions) in the classical limit~\cite{Berry85,SR2001,S2002,Haake2004,Haake2005,Saito}.

For quantum many-body systems with no well defined classical limit the quantum chaos conjecture can be taken as a definition of quantum chaos. Indeed, an extensive number of numerical studies (see, e.g., Refs.~\cite{ergodic1,ergodic2,ergodic3,ergodic4}) established that systems with random-matrix spectral correlations display many features that are intuitively connected to chaos. In particular, spectral correlations are a widespread diagnostic tool to test numerically whether a many-body system is expected to be ergodic. 
Until recently, however, the theoretical explanations of this phenomenon where extremely scarce: no analytical method was known to deduce the spectral correlations from the Hamiltonian of the system or from the time evolution operator. 

The situation has changed drastically over the last few years, when a number of settings and methods have been proposed to derive analytically the \emph{spectral form factor} (SFF) (i.e.\ the Fourier transform of the two-point correlation function of energy levels). Specifically, Refs.~\cite{KLP, RP20} established random matrix spectral fluctuations in long-ranged (but non-mean-field) periodically driven spin chains. Further on, Refs.~\cite{Chalker,Chalker2, Chalker3, Chalker4} demonstrated the emergence of random-matrix spectral correlations in periodically driven local random circuits, where the interactions are determined by random two-site gates acting on neighbouring sites and chosen (once and for all) at the beginning of the evolution. In particular, analytical results were provided in the limit of large local Hilbert space dimension. Finally, Ref.~\cite{BKP:SFF} provided an exact result for the spectral form factor in the self-dual kicked Ising model: a system of spin-1/2 variables which are interacting locally with an Ising Hamiltonian and are periodically ``kicked" by a longitudinal magnetic field. The term ``self-dual" indicates that the longitudinal field and the Ising coupling are set to specific values. The key property to obtain the exact result is that, at aforementioned specific values of the couplings, the problem can be formulated in terms of a transfer matrix ``in space" (i.e.\ propagating in the spatial direction, rather than in the temporal one) which is \emph{unitary}.

The spectral form factor alone, however, is not a sufficient evidence for claiming the chaoticity of a system. Indeed, to invoke the quantum chaos conjecture one needs all the spectral correlation functions, not just the two point one. The goal of this paper is to provide such a result in the case of the self-dual kicked Ising model. We will generalise the space-transfer-matrix method of Ref.~\cite{BKP:SFF} to find expressions for higher moments of the spectral form factor and use them to obtain rigorous lower bounds. Then, we will demonstrate numerically that the bounds are saturated.  

The rest of the paper is laid out as follows. In Sec.~\ref{sec:model} we introduce the model and the quantities of interest (i.e. the spectral form factor and its higher moments). In Sec.~\ref{sec:RMT} we identify the ensembles of random matrices which is relevant for the self-dual kicked Ising model and provide a prediction for the higher moments of the spectral form factor. In Sec.~\ref{sec:Lowerbound} we provide the aforementioned lower bounds on the higher moments and in Sec.~\ref{sec:montecarlo} we show numerically that the bounds are saturated. Finally, Sec.~\ref{sec:conclusions} contains our conclusions. Appendix~\ref{sec:app} reports some details on the spectrum of the space transfer matrix for short (finite) times.

\section{The model}
\label{sec:model}

We consider the \emph{self-dual} kicked Ising model~\cite{BKP:SFF, Guhr:duality}, described by the following time-dependent Hamiltonian 
\be
H_{\rm KI}[\boldsymbol h;t]= H_{\rm I}[\boldsymbol h]+ \delta_p(t) H_{\rm K}\,,
\label{eq:ham}
\ee
where $\delta_p(t)=\sum_{m=-\infty}^\infty\delta(t- m \tau )$ is the periodic delta function and 
\begin{align}
&H_{\rm I}[\boldsymbol h]\equiv \frac{\pi}{4\tau} \sum_{j=1}^{L} (\sigma^z_j \sigma^z_{j+1}-\1_L)+ \frac{\pi}{4\tau} \sum_{j=1}^{L} h_j \sigma^z_{j}\,,\label{eq:classicalising}\\
&H_{\rm K} \equiv \frac{\pi}{4\tau} \sum_{j=1}^{L} \sigma^x_j \,.\label{eq:kick}
\end{align}
Here $\tau$ is time interval between two kicks, $L$ denotes the volume of the system, $\1_x$ is the identity operator in $({\mathbb C}^2)^{\otimes x}$, $\{\sigma_j^{\alpha}\}_{\alpha=x,y,z}$ are Pauli matrices at position $j$, and we impose 
\be
\sigma_{L+1}^{\alpha}=\sigma_1^{\alpha}.
\ee
The parameter $\boldsymbol h=(h_1,\ldots, h_L)$ describes a position dependent longitudinal field measured in units of $\tau^{-1}$. From now on $\tau$ is set to 1 to simplify the notation.

The Floquet operator generated by \eqref{eq:ham} reads as 
\be
\hspace{-0.2cm}U_{\rm KI}[\boldsymbol h]= 
\!T\exp\!\!\left[-i\!\int_0^1\!\!\!{\rm d}s\, H_{\rm KI}[\boldsymbol h;s]\right]\!\!=\!
e^{-i H_{\rm K}}e^{-i H_{\rm I}[\boldsymbol h]}\,.
\label{eq:floquet}
\ee
In Floquet systems it is customary to introduce \emph{quasienergies} $\{\varphi_n\}$ defined as the phases of the eigenvalues of the Floquet operator. The quasienergies take values in the interval 
$[0,2\pi]$ and their number $\mathcal N = 2^L$ is the dimension of the Hilbert space where \eqref{eq:ham} acts, namely 
\be
{\cal H}_L=({\mathbb C}^2)^{\otimes L}\,.
\ee
To characterise the distribution of quasienergies (and especially the correlations among them) it is convenient to consider the SFF
\be
K(t,L)\equiv |{\rm tr} [U^t_{\rm KI}[\boldsymbol h]]|^2\,.
\label{eq:SFF}
\ee
This quantity represents an efficient diagnostic tool able to tell apart chaotic (non-integrable) systems from integrable ones even in the thermodynamic limit ($L\to\infty$). Indeed, the former are believed to show uncorrelated (Poisson distributed) quasienergies~\cite{BT} while the latter to display quasienergies distributed as in random unitary matrices~\cite{ergodic1, ergodic2, ergodic3, ergodic4, KLP, RP20, Chalker,Chalker2, Chalker3, Chalker4, BKP:SFF}. In the first case the SFF~\eqref{eq:SFF} is independent of time, while it shows a linear ramp in the second.

Importantly, the probability distribution of the SFF \emph{does not} become a delta function in the thermodynamic limit~\cite{nonSA,COE:SFF} (this property is commonly referred to as ``non-self-averaging" property~\cite{nonSA}). This means that, to have a meaningful comparison with the prediction of RMT, one has to study the probability distribution of the SFF over an ensemble of systems. The ensemble can be formed by considering similar systems with different numerical values of the parameters or the same system at different times. Here we follow Ref.~\cite{BKP:SFF} and consider the distribution of \eqref{eq:SFF} in an ensemble formed by self-dual kicked Ising models \eqref{eq:ham} with random longitudinal fields. Specifically we assume that the longitudinal magnetic fields at different spatial points $h_j$ are independently distributed Gaussian variables with mean value $\bar h$ and variance $\sigma^2 > 0$. 
Differently from Ref.~\cite{BKP:SFF}, however, here we are interested in the thermodynamic limit of all moments of the distribution of $|{\rm tr} [U^t_{\rm KI}[\boldsymbol h]]|^2$ not just in the average. Namely we consider  
\begin{align}
\!\!K_n(t)\equiv \lim_{L\to\infty}\mathbb E_{\boldsymbol h}\left[|{\rm tr} [U^t_{\rm KI}[\boldsymbol h]]|^{2n}\right]\!,\qquad n\geq 1\,,
\label{eq:Kmoments}
\end{align}
where the symbol $\mathbb E_{\boldsymbol h}[\cdot]$ denotes the average over the longitudinal fields 
\be
\mathbb E_{\boldsymbol h}\left[ f(\boldsymbol h)\right] = \int_{-\infty}^\infty f(\boldsymbol h) \prod_{j=1}^L e^{-{(h_{j}-\bar h)^2}/{2\sigma^2}}\frac{{\rm d}h_j}{\sqrt{2\pi}\sigma}\,.
\label{eq:average}
\ee
In this language the thermodynamic limit of the SFF corresponds to $K_1(t)$.

\section{Prediction of Random Matrix Theory}
\label{sec:RMT}

Before computing \eqref{eq:Kmoments} in the self-dual kicked Ising model we compute the moments for an ensemble of \emph{random unitary matrices} subject to the same constraints --- or symmetries --- as the Floquet operator $U_{\rm KI}[\boldsymbol h]$ (cf. Eq.~\eqref{eq:floquet}). Indeed, due to some special symmetries of $U_{\rm KI}[\boldsymbol h]$, such ensemble is not the ``standard" circular orthogonal ensemble (COE) --- composed of symmetric unitary matrices. To see that let us start by reviewing the symmetries of $U_{\rm KI}[\boldsymbol h]$. 

\subsection{Symmetries of the Time-Evolution Operator}

To analyse the symmetries of \eqref{eq:floquet} it is convenient to make the following basis transformation
\be
U_{\rm KI}[\boldsymbol h]\mapsto e^{-i H_{\rm K}/2}U_{\rm KI}[\boldsymbol h] e^{i H_{\rm K}/2}\equiv \bar U_{\rm KI}[\boldsymbol h].
\label{eq:transformation}
\ee 
This transformation leaves \eqref{eq:Kmoments} invariant and brings the operator in a manifestly symmetric form 
\be
\bar U_{\rm KI}[\boldsymbol h]=\bar U_{\rm KI}^T[\boldsymbol h].
\label{eq:sym1}
\ee
Since $\bar U_{\rm KI}[\boldsymbol h]$ is unitary and symmetric we immediately have 
\be
C^{\dag} \bar U^{\phantom{-1}}_{\rm KI}[\boldsymbol h] C= \bar U_{\rm KI}^*[\boldsymbol h] = \bar U_{\rm KI}^{-1}[\boldsymbol h]\,,
\ee
where ${(\cdot)}^*$ denotes complex conjugation in the \emph{computational basis} (the standard Pauli basis where both matrices $\sigma^x$ and $\sigma^z$ are real) and $C$ is the anti-unitary operator implementing it in the Hilbert space. This is the most obvious anti-unitary symmetry of the time-evolution operator and corresponds to the standard time-reversal symmetry $T$ (with $T^2=\1$).

As observed in Ref.~\cite{BWAGG:SFF}, however, $T$ is not the only anti-unitary symmetry of $\bar U_{\rm KI}[\boldsymbol h]$. Indeed, defining    
\begin{align}
&F_y\equiv \prod_{j=1}^L \sigma^y_j = (\sigma^y)^{\otimes L} = F_y^\dag=F_y^{-1},\label{eq:FY}\\
&U\equiv  \exp\!\!\left[i \frac{\pi}{4} \sum_{j=1}^{L}  (\sigma^z_j\sigma^z_{j+1}-\1_L)\right],\label{eq:U}
\end{align}
and noting
\begin{align}
F_y \sigma^{x,z}_{j} F_y^\dag & =  - \sigma^{x,z}_{j},\\
U^2 &= \1_L,
\end{align}
one readily finds 
 \be
F_y^{\dag} \bar U^{\phantom{-1}}_{\rm KI}[\boldsymbol h] F_y= \bar U_{\rm KI}^*[\boldsymbol h] = \bar U_{\rm KI}^{-1}[\boldsymbol h]\,.
\label{eq:sym2}
\ee
This equation shows that $\bar U^{\phantom{-1}}_{\rm KI}[\boldsymbol h]$ and $\bar U^{\ast}_{\rm KI}[\boldsymbol h]$ are related by a similarity transformation and, therefore, it implies that the spectrum of $\bar U^{\phantom{-1}}_{\rm KI}[\boldsymbol h]$ is symmetric around the real axis, i.e. ${{\rm sp}(\bar U_{\rm KI}[\boldsymbol h])={\rm sp}(\bar U^*_{\rm KI}[\boldsymbol h])={\rm sp}(\bar U_{\rm KI}[\boldsymbol h])^*}$, i.e. all quasienergies form pairs $\{\varphi_n,-\varphi_n\}$.

Reshaping \eqref{eq:sym1} and \eqref{eq:sym2} we will now see that they correspond to the constraints on random matrix ensembles associated to two compact symmetric spaces~\cite{Duenez, Duenezthesis} (two different symmetric spaces will correspond to even and odd $L$). To see this, we note that, permuting the computational basis, $F_y$ can be brought to one of the two following block-diagonal forms, depending on the parity of $L$ 
\begin{align}
& P_1 F_y P_1^T = 
\begin{bmatrix}
\sigma^y & 0 & \cdots & 0 \\ 
0 &\sigma^y  & \cdots & 0 \\ 
\vdots & \vdots & \ddots & \vdots\\
0 &  0 & \cdots  & \sigma^y  
\end{bmatrix},
& L \quad \textrm{odd},\label{eq:Fyodd}\\
\notag\\
& P_2 F_y P_2^T = 
\begin{bmatrix}
s_1 \sigma^x & 0 & \cdots & 0 \\ 
0 & s_2 \sigma^x  & \cdots & 0 \\ 
\vdots & \vdots & \ddots & \vdots\\
0 & 0 & \cdots &  s_{\mathcal N'} \sigma^x  
\end{bmatrix},
& L \quad \textrm{even},\label{eq:Fyeven}
\end{align}
where $P_1 P_1^T = P_2 P_2^T = \1_{\mathcal N}$, $\mathcal N' = \mathcal N/2$, and $\{s_j\}_{j=1}^{\mathcal N'}$ is a specific string of $+1$s and $-1$s. 

\subsubsection{$L$ odd}

The matrix \eqref{eq:Fyodd} is a non-singular real skew-symmetric matrix (i.e. $\Omega_{\mathcal N'}=-\Omega^T_{\mathcal N'}$) multiplied by $i^L$, this means that defining  
\be
\hat U_{\rm KI}[\boldsymbol h]  \equiv P_1 \bar U_{\rm KI}[\boldsymbol h] P_1^T, \qquad  L \quad \textrm{odd}, 
\ee
we have that \eqref{eq:sym1} and \eqref{eq:sym2} become 
\begin{align}
\hat U_{\rm KI}[\boldsymbol h] &=\hat U_{\rm KI}^T[\boldsymbol h],\\
\hat U^{-1}_{\rm KI}[\boldsymbol h]  &=\Omega^T_{\mathcal N'} \hat U_{\rm KI}^{T}[\boldsymbol h] \Omega_{\mathcal N'}.
\end{align}
Namely the unitary matrix $\hat U_{\rm KI}[\boldsymbol h]$ is constrained to be \emph{symmetric} and \emph{symplectic}. The compact symmetric space characterised by this constraint is 
\be
S_{-}(\mathcal N')\equiv Sp(\mathcal N')/U(\mathcal N')
\ee
and corresponds to CI in Cartan's classification~\cite{Duenez, Duenezthesis}. Note that here we denoted by ${Sp(N)\subset U(2N)}$ the unitary-symplectic group of ${2N\times 2N}$ unitary matrices $m$ fulfilling ${m^{-1}\!=\Omega_{N}^T\!\cdot\! m^T\! \cdot\! \Omega_{N}}$, sometimes denoted also by $U\!Sp(2N)$.

As shown in~\cite{Duenez, Duenezthesis} matrices belonging to this symmetric space can be parametrized by 
\be
\!\!g \begin{bmatrix}
\1_{L-1} & 0 \\ 
0 &-\1_{L-1}
\end{bmatrix}\! g^{-1}\!\! \begin{bmatrix}
\1_{L-1} & 0 \\ 
0 &-\1_{L-1}
\end{bmatrix},\,\, g\in Sp(\mathcal N').
\ee  

\subsubsection{$L$ even}

The matrix on the r.h.s. of \eqref{eq:Fyeven}, instead, can be written as the square of 
\be
S = e^{-i \tfrac \pi 4}
\begin{bmatrix}
e^{i \pi s_1 \sigma^x/4} & 0 & \cdots & 0 \\ 
0 & e^{i \pi s_2 \sigma^x/4}  & \cdots & 0 \\ 
\vdots & \vdots & \ddots & \vdots\\
0 & 0 & \cdots &  e^{i \pi s_{\mathcal N'} \sigma^x/4}  
\end{bmatrix}.
\ee
Moreover, it can be brought to the following diagonal form by means of an orthogonal transformation $P_3$ 
\be
P_3 S^2 P_3^T = 
\begin{bmatrix}
\1_{L-1} & 0 \\ 
0 &-\1_{L-1}
\end{bmatrix}\,.
\ee
So that defining 
\be
\hat U_{\rm KI}[\boldsymbol h]  \equiv P_3 S^* P_2 \bar U_{\rm KI}[\boldsymbol h] P_2^T S P_3^T, \qquad  L \quad \textrm{even}, 
\ee
we have 
\begin{align}
\hat U_{\rm KI}[\boldsymbol h] &=\hat U_{\rm KI}^*[\boldsymbol h],\\
\hat U_{\rm KI}[\boldsymbol h]  &= \begin{bmatrix}
\1_{L-1} & 0 \\ 
0 &-\1_{L-1}
\end{bmatrix} \hat U_{\rm KI}^{T}[\boldsymbol h] \begin{bmatrix}
\1_{L-1} & 0 \\ 
0 &-\1_{L-1}
\end{bmatrix}.
\label{eq:condeven2}
\end{align}
This means that the unitary matrix $\hat U_{\rm KI}[\boldsymbol h]$ is constrained to be \emph{real orthogonal} and fulfil \eqref{eq:condeven2}. The compact symmetric space characterised by these constraints is 
\be
S_{+}(\mathcal N')\equiv O(2 \mathcal N')/{(O(\mathcal N')\!\times\! O(\mathcal N'))}
\ee 
and corresponds to BDI in Cartan's classification~\cite{Duenez, Duenezthesis}. As shown in~\cite{Duenez, Duenezthesis} the matrices in this symmetric space can be parametrized by 
\be
\!\!\! g \begin{bmatrix}
\1_{L-1} & 0 \\ 
0 &-\1_{L-1}
\end{bmatrix}\! g^{-1}\!\! \begin{bmatrix}
\1_{L-1} & 0 \\ 
0 &-\1_{L-1}
\end{bmatrix},\,\, g\in O(2 \mathcal N').
\ee  

\subsection{Relevant Random-Matrix Ensembles}

The random matrix ensembles corresponding to the symmetric spaces $S_-(\mathcal N')$ and $S_+(\mathcal N')$ have been introduced in Refs.~\cite{Duenez, Duenezthesis}: for both ensembles one finds that the quasienergies come in pairs of opposite values $\{\varphi_j,-\varphi_j\}_{j=1}^{\mathcal N'}$. Moreover, from the probability measure induced by the Riemannian metric of the symmetric spaces one finds the following (joint) probability distributions for  $\boldsymbol \varphi=\{\varphi_j\}_{j=1}^{\mathcal N'}\in [0,\pi]^{\mathcal N'}$
\begin{align}
P_{-}(\boldsymbol \varphi) &\propto \!\!\!\!\!\!\!\!\prod_{1\leq i<j\leq \mathcal N'} \!\!\!\!\!\!\!|\cos \varphi_i\! - \!\cos \varphi_j| \prod_{i=1}^{ \mathcal N'} \sin\varphi_i, \label{eq:PCI}\\
P_{+}(\boldsymbol \varphi) &\propto \!\!\!\!\!\!\!\!\prod_{1\leq i<j\leq \mathcal N'} \!\!\!\!\!\!\!|\cos \varphi_i\! - \!\cos \varphi_j|,\label{eq:PBDI} 
\end{align}
where the proportionality constant is chosen to ensure that their integral over $[0,\pi]^{\mathcal N'}$ is one. Changing variables from $\varphi_j$ to ${x_j=\cos\varphi_j \in [-1,1]}$, both \eqref{eq:PCI} and \eqref{eq:PBDI} are brought into the so-called Jacobi-ensemble form~\cite{ForresterBook}
\be
P_{ab}^J(\boldsymbol x) \propto \!\!\!\!\!\!\!\!\prod_{1\leq i<j\leq \mathcal N'} \!\!\!\!\!\!\!|x_i - x_j|^\beta \prod_{i=1}^{ \mathcal N'} (1-x_i)^{a\beta/2}(1+x_i)^{b\beta/2}\!,
\label{eq:JacobiPD}
\ee
with ${\beta=1}$ and, respectively, ${a=b=0}$ for $S_-(\mathcal N')$ and ${a=b=-1}$ for $S_+(\mathcal N')$.

\subsection{Thermodynamic Limit of the Moments}

Let us now turn to the main objective of this section: computing the moments of SFF \eqref{eq:Kmoments} where the matrix $U_{\rm KI}[\boldsymbol h]$ is replaced by a random matrix in $U \in S_{\pm}(\mathcal N')$ and the average $\mathbb E_{\boldsymbol h}[\cdot]$ is replaced by  
\be
\mathbb E^{\pm}_{\boldsymbol \varphi}\left[ f \right] = \!\!\!\!\int\limits_{[0,\pi]^{\mathcal N'}} \!\!\!\!\!\!f(\boldsymbol \varphi) P_{\pm}(\boldsymbol \varphi) \prod_{j=1}^{\mathcal N'} {\rm d}\varphi_j\,,
\ee
where $+$/$-$ are respectively chosen for $L$ even/odd. To compute the moments \eqref{eq:Kmoments} it is convenient to find the full probability distribution of the \emph{linear statistics}  
\be
T_{t,L} = {\rm tr}[U^t] = \sum_{j=1}^{\mathcal N} e^{i t \varphi_j} = 2 \sum_{j=1}^{\mathcal N'} \cos({t \varphi_j})\,,
\label{eq:definitionT}
\ee
where $ \varphi_j$ are the quasienergies of $U$ and in the last step we used that they can only appear in complex conjugated pairs. An immediate consequence of this relation is that, as opposed to what happens in the COE, the random variable $T_{t,L}$ is real. 

Let us start by considering the average of \eqref{eq:definitionT}. First we note that, introducing the $n$-point function of the density of quasienergies 
\be
\!\!\!\rho_{\pm,n}(x_1,\ldots,x_n)\!= 
\mathbb E^{\pm}_{\boldsymbol \varphi}\!\!\left[  \sum_{j_1\neq\ldots \neq j_n=1}^{\mathcal N} \prod_{k=1}^n \!\delta(x_k-\varphi_{j_k})\!\right]\!\!,
\ee
the average can be expressed as  
\be
\mathbb E^{\pm}_{\boldsymbol \varphi}\left[ T_{t,L} \right] = \int {\rm d} \varphi\,\, 2 \cos(\varphi t) \rho_{\pm, 1}(\varphi).
\ee
Since we are interested in the thermodynamic limit ($L\to\infty$) we do not need to find the statistics of $T_{t,L}$ exactly: it is sufficient to find its leading behaviour for large $\mathcal N'=2^{L-1}$. This can be efficiently done using ``log-gas methods"~\cite{Dyson, Mehtabook, ForresterBook}, i.e. studying the statistical mechanics of quasienergies through the formal analogy with a  gas of charged particles in two dimensions (confined in a one-dimensional domain). Specifically, here we will follow the treatment of Ref.~\cite{ForresterBook}. 

{First we observe that the probability distribution \eqref{eq:JacobiPD} of the Jacobi ensemble is equivalent to the Boltzmann factor of a one-component log-potential gas confined to the interval $[-1,1]$ with particles of unit charge at positions $\{x_1,\ldots,x_{\mathcal N'}\}$ and a neutralising background charge density  
\begin{align}
&\rho_{b}(x)=  -\frac{{\mathcal N'}+2/\beta-1+(a+b)/2}{\pi (1-x^2)^{1/2}} \notag\\
&\qquad+\!\! \left[\frac{a-1}{2}\!+\!\frac{1}{\beta}\right]\!\delta(x-1)\!+\!\! \left[\frac{b-1}{2}\!+\!\frac{1}{\beta}\right]\!\delta(x+1)\,,
\label{eq:rhob}
\end{align}
where we neglected $O(1/{\mathcal N'})$. This statement is proven in Proposition 3.6.3 of Ref.~\cite{ForresterBook} (see also Exercises 14.2). Eq.~\eqref{eq:rhob} can be used to fix the density of the gas by requiring that, in the thermodynamic limit, the system is locally neutral, so that 
\be
\lim_{{\mathcal N'}\to \infty} \rho_{b}(x)+\rho_{\pm, 1}(x)=0\,.
\ee
In particular, changing variables to $\varphi = {\cos}^{-1} x$ and setting $\beta=1$ and ${a=b=1\,(0)}$ for $U \in S_{+(-)}(\mathcal N')$ we find 
\be
\!\!\!\!\!\lim_{L\to\infty} \rho_{\pm, 1}(\varphi)-\frac{\mathcal N'}{\pi}= \pm \left[\frac{1}{2}\delta(\varphi)+\frac{1}{2}\delta(\varphi-\pi)-\frac{1}{2\pi}\right]\!.
\ee
This result agrees with the infinite $L$ limit of the exact one-point function in the Jacobi ensemble (\emph{cf}. Proposition 6.3.3 of Ref.~\cite{ForresterBook}). Moreover, it also implies}  
{
\begin{align}
\lim_{L\to\infty} \mathbb E^{\pm}_{\boldsymbol \varphi}\left[ T_{t,L} \right]  &= \pm \frac{1+(-1)^t}{2}= \pm {\rm mod}(t+1,2)\,,
\label{eq:predictiontrace}
\end{align}}
where ${\rm mod}(n,m)= n\, {\rm mod}\, m$ is the mod function. 

Next, we consider the variance 
\bw
\begin{align}
{\rm Var}_\pm(T_{t,L}) & \equiv  \mathbb E^{\pm}_{\boldsymbol \varphi}\left[ T^2_{t,L} \right] -  \mathbb E^{\pm}_{\boldsymbol \varphi}\left[ T_{t,L} \right]^2 = 4 \int\limits_{[0,\pi]^{2}} \!\!{\rm d} \varphi_1 {\rm d} \varphi_2\,\,  \cos(\varphi_1 t)\cos(\varphi_2 t)  \left(\rho^c_{\pm, 2}(\varphi_1,\varphi_2)+ \rho_{\pm,1}(\varphi_1)\delta(\varphi_1-\varphi_2)\right),
\label{eq:variance}
\end{align}
\ew
where $\rho^c_{\pm, 2}(\varphi_1,\varphi_2)$ is the connected two point function 
\be
\rho^c_{\pm, 2}(\varphi_1,\varphi_2)\equiv \rho_{\pm, 2}(\varphi_1,\varphi_2)-\rho_{\pm, 1}(\varphi_1)\rho_{\pm, 1}(\varphi_2).
\ee
{The large-$L$ behaviour of the quantity 
\be
K(\varphi_1,\varphi_2) \equiv \rho^c_{\pm, 2}(\varphi_1,\varphi_2)+ \rho_{\pm,1}(\varphi_1)\delta(\varphi_1-\varphi_2)
\ee
can again be computed in the log-gas framework. In this case one uses a linear response argument (see Chapter 14.3 of Ref.~\cite{ForresterBook}). In essence one imagines to add an infinitesimal charge $\delta q$ to the log-gas system, which is assumed to behave like a perfect conductor. Therefore, the charges of the log gas redistribute to screen $\delta q$. In this setting one can show that $K(\varphi_1,\varphi_2)$ is proportional to the to crossed derivative (in both $\varphi_1$ and $\varphi_2$) of the electronic potential created by the displaced charges. In the case of the Jacobi ensemble this leads to 
\bw
\be
\lim_{L\to\infty}  K(\varphi_1,\varphi_2) = -\frac{1}{\beta \pi^2} \frac{1}{\sin\varphi_1} \frac{\partial^2}{\partial \varphi_1\partial \varphi_2} \sin\varphi_2 \log|\cos\varphi_1-\cos\varphi_2|\,.
\ee
Substituting in \eqref{eq:variance} one finds 
\be
\lim_{L\to\infty} {\rm Var}_\pm(T_{t,L}) = \frac{4 t}{\beta \pi^2} \int\limits_{[0,\pi]} \!\!{\rm d} \varphi_1 \cos(\varphi_1 t) \int\limits_{[0,\pi]}\!\!{\rm d} \varphi_2\,\,  \frac{\sin(\varphi_2 t) \sin\varphi_2}{\cos\varphi_1-\cos\varphi_2}\,.
\ee
\ew
Note that, with the change of variables $\cos\varphi_1 \to x$ and $\cos\varphi_2 \to y$, this expression corresponds to Eq.~14.56 of Ref.~\cite{ForresterBook} with $a(\cos\theta)=2\cos(t\theta)$ and $\beta=1$. Carrying on the integrals we find
\be
\lim_{L\to\infty}{\rm Var}^\pm(T_{t,L}) = 2t\,. 
\ee
Using the results for mean and variance we can now deduce the large $L$ limit of the full probability distribution of $T_{t,L}$, namely   
\be
P_{\pm,T}(x) \equiv \!\!\!\!\int\limits_{[0,\pi]^{\mathcal N'}} \delta(x-T_{t,L})\, P_{\pm}(\boldsymbol \varphi) \prod_{j=1}^{\mathcal N'} {\rm d}\varphi_j\,.
\ee
Indeed, using again a linear response argument (see Chapter 14.4 of Ref.~\cite{ForresterBook}), one can show that in this limit $P_{\pm,T}(x)$ becomes Gaussian (see Eq.~14.68 of Ref.~\cite{ForresterBook}) so that we finally obtain 
\be
\lim_{L\to\infty}P_{\pm,T}(x) = \frac{1}{\sqrt{4\pi t}} e^{-\displaystyle \frac{(x \mp {\rm mod}(t+1,2))^2}{4t}}.
\label{eq:probdistJacobi}
\ee}
The probability distribution \eqref{eq:probdistJacobi} produces the following central moments in the thermodynamic limit 
\begin{align}
\!\!\!\!\!C_n(t)\!=\!\!\!\lim_{L\to\infty}\!\!\mathbb E^{\pm}_{\boldsymbol \varphi}\!\left[|T_{t,L} \!-\! \mathbb E^{\pm}_{\boldsymbol \varphi}[T_{t,L}]|^{2n}\right]\!\!=\! (2t)^n (2n-1)!!,
\end{align}
and therefore \eqref{eq:Kmoments} read as  
\be
K_n(t) = 
 \begin{cases}
\displaystyle \sum_{k=0}^n \binom{2n}{2k} C_k(t) & t\,\,\text{even}\\
\\
\displaystyle C_n(t) \,. & t\,\,\text{odd}
\end{cases}\,.
\label{eq:cumulantsJacobi}
\ee
The result \eqref{eq:cumulantsJacobi} is very different from the one found for $U\in {\rm COE}$. Indeed, in the latter case the expression \eqref{eq:definitionT} is complex and Ref.~\cite{COE:SFF} found the following joint distribution for its real and imaginary part (respectively $x$ and $y$) in the thermodynamic limit
\be
\lim_{L\to\infty}P_{T}(x,y) = \frac{1}{2\pi t} e^{-\displaystyle \frac{x^2+y^2}{2t}}.
\label{eq:probdistCOE}
\ee
This distribution yields
\be
K_n(t) = (2t)^n n!\,.
\label{eq:cumulantsCOE}
\ee
We see that, even though \eqref{eq:cumulantsCOE} and \eqref{eq:cumulantsJacobi} agree for $n=1$ and $t$ odd, they are generically very different. In particular the moments \eqref{eq:cumulantsJacobi} are much larger that \eqref{eq:cumulantsCOE} indicating  that the fluctuations in the ensembles $S_{\pm}(\mathcal N')$ are larger than those in the COE. 

\section{Lower Bound from the space-transfer-matrix approach}
\label{sec:Lowerbound}

Equipped with the random matrix theory prediction~\eqref{eq:cumulantsJacobi} we can now move on to our main goal: computing the moments ${K}_n(t)$ in the self-dual kicked Ising model. In this section we will determine a rigorous lower bound for ${K}_n(t)$.

\subsection{Transfer Matrix in Space}

To derive the lower bound we will follow Ref.~\cite{BKP:SFF} and use the transfer matrix \emph{in space}. The starting point is the following identity, which holds for the self dual kicked Ising model~\cite{BKP:SFF,Guhr:duality}  
\be
{\rm tr}\!\left[ U_{\rm KI}[\boldsymbol{h}]^t\right]={\rm tr}\!\left(\prod_{j=1}^{L}\tilde U_{\rm KI}[h_j\boldsymbol{ \varepsilon}]\right)\,.
\label{eq:duality}
\ee
Here $\boldsymbol{\varepsilon}$ is a vector with $t$ entries equal to one and $\tilde  U_{\rm KI}[\boldsymbol h]$ takes the form \eqref{eq:floquet} with the only difference that the size $L$ is replaced by $t$ in \eqref{eq:classicalising} and \eqref{eq:kick}. Note that the trace on the right hand side of Eq.~\eqref{eq:duality} is over $\mathcal H_t= (\mathbb C^2)^{\otimes t}$. 

\begin{figure}
\includegraphics[width=0.4\textwidth]{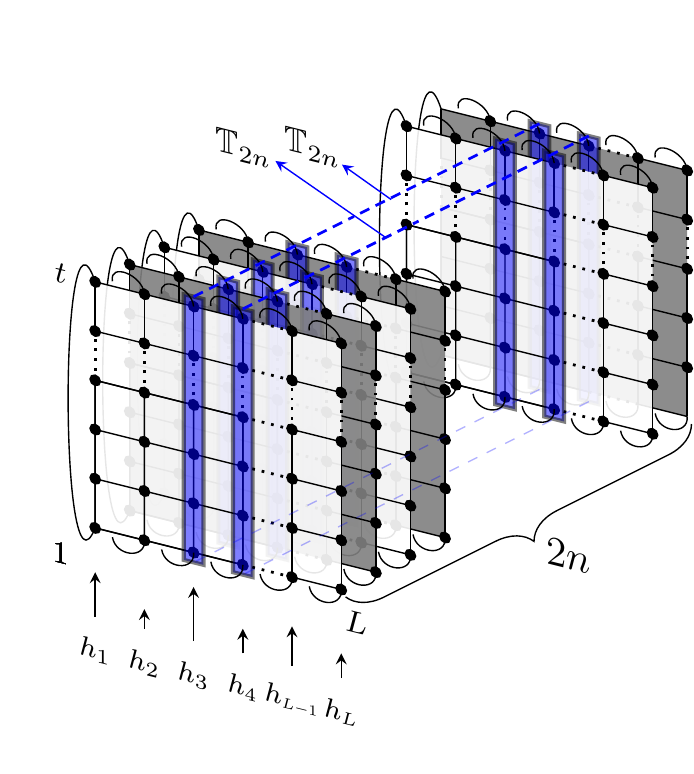}
\caption{An illustration of the $n$-th moment of the spectral form factor $K_{n}(t)$. The lattice depicts a system of $L$ spins that are propagated to time $t$. $\mathbb{T}_{2n}$ acts as a transfer matrix on $2n$ copies of the lattice. The average ($\mathbb{E}_{h_j}$) is performed over the longitudinal magnetic fields $h_j$. The loops on the edges of the lattice indicate that we need to compute the trace of $\mathbb{T}_{2n}$ to get the $n$-th moment of the spectral form factor.}
\label{figure1}
\end{figure}

Equation~\eqref{eq:duality} can be used to rewrite the $n$-th moment of the SFF as follows 
\be
{K}_n(t)=\lim_{L\to \infty}{\rm tr} \left(\mathbb{T}^L_{2n}\right)\,,
\label{eq:traceT}
\ee
with $\mathbb{T}_{2n}\in \text{End}(\mathcal{H}_t^{\otimes 2n})$ defined as 
\be
\mathbb{T}_{2n}=\mathbb E_{\boldsymbol h}\left[ \left(\tilde{U}_{\rm KI}\left[h_j\boldsymbol{ \varepsilon}\right]\otimes \tilde{U}_{\rm KI}^{*}\left[h_j\boldsymbol{ \varepsilon} \right]\right)^{\otimes n}\right]\,.
\label{eq:transfer matrix}
\ee
By looking at the graphical representation in Fig.~\ref{figure1} we see that $\mathbb{T}_{2n}$ plays the role of a space transfer matrix on a multi-sheeted two dimensional lattice. 
The simplification in Eq.~\eqref{eq:traceT} is possible because the matrices $U_{\rm KI}[h_j\boldsymbol{ \varepsilon}]$ on the r.h.s.\ of Eq.~\eqref{eq:duality} depend on longitudinal magnetic fields at different positions (which we assumed to be independently distributed) and the average factorises. Moreover, the Gaussian integral can be computed analytically yielding
\be
\mathbb{T}_{2n}=\tilde{\mathcal{U}}_{{\rm KI}, n}^{\phantom{*}}\otimes\tilde{\mathcal{U}}_{{\rm KI}, n}^*\cdot \mathbb{O}_{n,n}\,,
\label{eq:T4}
\ee
where we introduced
\begin{align}
\mathbb{O}_{n,m}&\equiv \exp\Bigl[-\frac{\sigma}{2}\bigl(\mathcal{M}_{\alpha, n}\otimes \1_{{tm}}-\1_{{tn}}\otimes\mathcal{M}_{\alpha, m}\bigr)^2\Bigr],
\label{eq:O2n}\\
\mathcal{M}_{\alpha, n}&\equiv\sum_{j=1}^n  \1_{t}^{\otimes(j-1)} \otimes M_\alpha \otimes \1_{t}^{\otimes(n-j)} ,\\
\tilde{\mathcal{U}}_{{\rm KI}, n}&\equiv (\tilde{U}_{\rm KI})^{\otimes n}\,.
\end{align}
Note that here  
\be
\tilde{U}_{\rm KI}\equiv\tilde{U}_{\rm KI}\left[\bar{h}\boldsymbol{ \varepsilon}\right],
\ee
is the transfer matrix in space at the average magnetic field, and 
\be
M_\alpha \equiv \sum_{\tau=1}^{t}\sigma_{\tau}^\alpha
\ee
is the magnetisation (in the $\alpha$ direction) for a chain of length $t$.

\subsection{Trace of ${U}_{\rm KI}^{t}[\boldsymbol h]$}
\label{Half moment}

Before embarking on the analysis of Eq.~\eqref{eq:traceT} it is useful to look at a simpler observable that can be studied with the same method, namely 
\be
B(t) \equiv  \lim_{L\to\infty} \mathbb E_{\boldsymbol h}\!\!\left[{\rm tr} [U^t_{\rm KI}[\boldsymbol h]]\right].
\label{eq:trUt}
\ee
Indeed, the RMT prediction for this quantity is non-trivial (cf.\ Eq.~\eqref{eq:predictiontrace}) and offers a convenient opportunity for testing the quantum chaos conjecture. Moreover, performing the calculation in this simple example will best illustrate some of the main ideas.

Considering \eqref{eq:trUt} and using \eqref{eq:duality} we have 
\be
B(t) =  \lim_{L\to\infty} {\rm tr}[\mathbb T^L] 
\label{eq:B}
\ee
where in this case the space-transfer matrix reads as 
\be
\mathbb T =\tilde{U}_{\rm KI}\exp\Bigl[{-\frac{\sigma}{2}M_z^2}\Bigr] \equiv \tilde{U}_{\rm KI}\, \mathbb{O}_{1,0}\,.
\label{eq:T1}
\ee

\begin{center}
\begin{figure}[t]
		\includegraphics[width=.85\textwidth]{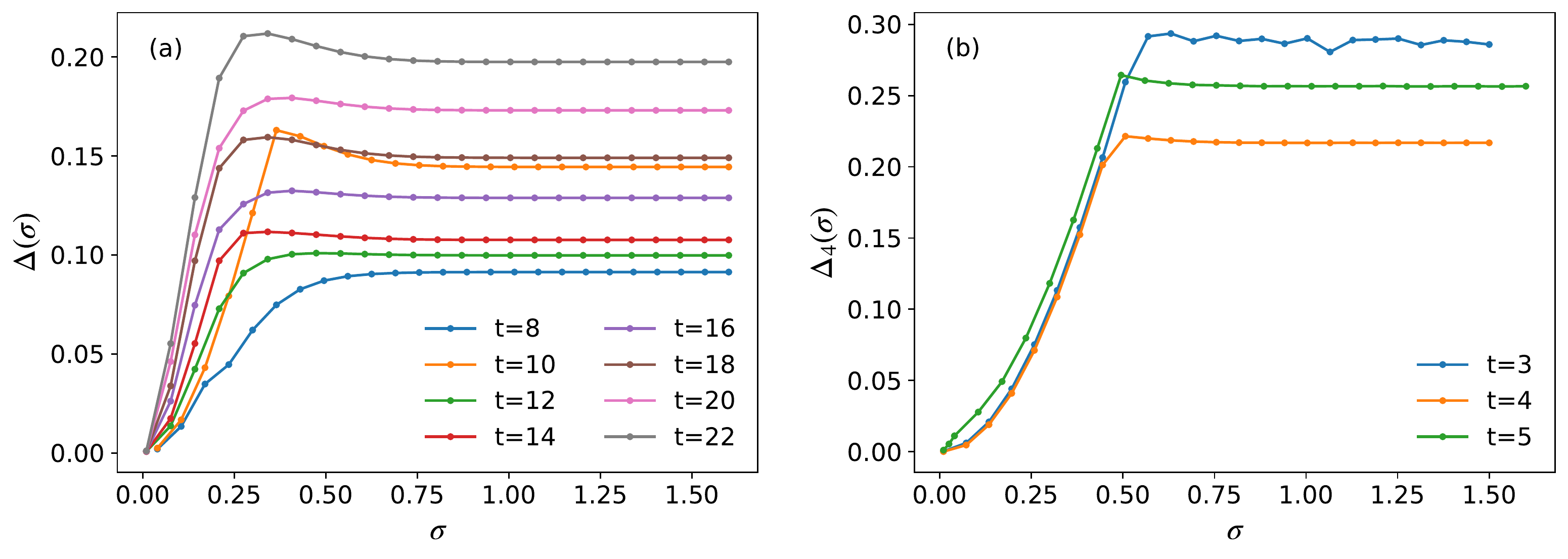}
	\caption{Spectral gap of transfer matrix $\mathbb{T}$, Eq.~(\ref{eq:T1}) (a), and transfer matrix $\mathbb{T}_4$, Eq.~(\ref{eq:T4}) (b), as a function of the disorder strength $\sigma$ for different times $t$. The average of the disorder is set to zero $\bar{h}=0$.}
	\label{figure_gap}	
\end{figure}
\end{center}

The limit \eqref{eq:B} can be computed as follows. First we observe that the eigenvalues of the transfer matrix $\mathbb T$ are at most of unit magnitude and, additionally, geometric and algebraic multiplicity of any eigenvalue with magnitude one coincide. This can be seen by using the relation 
\be
\mathbb T^{\dagger}\mathbb T=\mathbb{O}_{1,0}^{\dagger}\tilde{U}_{\rm KI}^{\dagger}\tilde{U}_{\rm KI}^{\phantom{\dag}}\mathbb{O}^{\phantom{\dag}}_{1,0}=\mathbb{O}^{\dagger}_{1,0}\mathbb{O}^{\phantom{\dag}}_{1,0}=\mathbb{O}^2_{1,0}\,,
\label{eq:property}
\ee
and reasoning as in the proof of Property 1 of Ref.~\cite{BKP:SFF}. {Moreover, following~\cite{BKP:SFF}, we assume that the spectral gap $\Delta=1-{\rm max}^{|\lambda|<1}_{\lambda \in {\rm Sp}(\mathbb T)} |\lambda|$ remains finite for all times (${\rm Sp}(A)$ denotes the spectrum of $A$). This is confirmed by exact diagonalisation of $\mathbb T$ for small times, see the left panel of Fig.~\ref{figure_gap}}. Putting all together we conclude that $B(t)$ is given by the number of eigenvectors $|A\rangle$ corresponding to unimodular eigenvalues. 
\onecolumngrid
\twocolumngrid
Next, we observe that --- because of Eq.~\eqref{eq:property} --- all unimodular eigenvalues of $\mathbb T$ lie in the eigenspace of $\mathbb{O}_{1,0}$ corresponding to eigenvalue one. Given the form of the operator $\mathbb{O}_{1,0}$, this means that all relevant eigenvectors $\ket{A}$ must be in the kernel of the operator $M_z$, i.e. 
\be
M_z|A\rangle=0\,.
\label{eq:kernel}
\ee
This relation allows us to conclude the analysis of odd times. Indeed, since in that case there can be no vectors in the kernel of $M_z$ (a spin-$1/2$ chain of odd length cannot have zero magnetisation), we find immediately find that $B(t)$ vanishes.  
 
To find the result for even $t$ we continue by acting on $|A\rangle$ with $\mathbb{T}$, this yields
\be
 \tilde{U}_{\rm KI}|A\rangle=e^{i\varphi}|A\rangle.
\label{eq:phase}
\ee
This equation, together with \eqref{eq:kernel}, implies
\begin{align}
M_{\alpha}|A\rangle &= 0, & &\alpha\in\{x, y, z\} 
\label{eq:condition1Tr}\\
\tilde U|A\rangle &=e^{i(\varphi+\tfrac \pi 4 t)}|A\rangle, & &\varphi\in[0,2\pi),
\label{eq:condition2Tr}
\end{align}
where $\tilde U$ is defined as in \eqref{eq:U} but with $L$ replaced by $t$. The first of these equations can be verified by using the identities 
\begin{align}
\tilde{U}_{\rm KI}M_z\tilde{U}_{KI}^{\dagger}&=- M_y,\\
e^{i\frac{\pi}{4}M_z}M_ye^{-i\frac{\pi}{4}M_z}&=M_x,
\end{align}
while the second follows from Eq.~\eqref{eq:condition1Tr} and \eqref{eq:phase}. 
 
Since the operator $\tilde U$ squares to $\1_{t}$ we have 
\be
e^{i\varphi}= \pm 1\,.
\label{eq:generaleigenvalueT}
\ee
A state that satisfies equations \eqref{eq:condition1Tr} and \eqref{eq:condition2Tr} is directly identified as 
\be
|\psi\rangle=\frac{1}{2^t}\prod_{\tau=1}^{t/2}(1-P_{\tau,\tau+t/2})|\uparrow\uparrow...\uparrow\downarrow\downarrow...\downarrow\rangle\,,
\label{eq:psi}
\ee
with $P_{i, j}=\frac{1}{2}\1+\frac{1}{2}\sum_{\alpha}\sigma^{\alpha}_i\sigma^{\alpha}_j$ being the transposition of the spins on sites $i$ and $j$. In particular, it is easy to verify that \eqref{eq:psi} fulfils \eqref{eq:condition1Tr} and \eqref{eq:condition2Tr} with 
\be
e^{i\varphi}= - 1\,.
\label{eq:eigenvalueT}
\ee
Assuming that \eqref{eq:psi} is the only eigenvector of $\mathbb T$ corresponding to unit magnitude eigenvalues we have 
{\be
B(t) = \begin{cases}
-{\rm mod}(t+1,2) & L\,\,\text{odd}\\
\phantom{-}{\rm mod}(t+1,2) & L\,\,\text{even}
\end{cases}\,,
\label{eq:traceKI}
\ee}
which agrees with the RMT prediction~\eqref{eq:predictiontrace}. Note that, for even values of $L$, Eq.~\eqref{eq:traceKI} gives a lower bound for $B(t)$. Indeed, given the general structure \eqref{eq:generaleigenvalueT} of the eigenvalues one can immediately see that the contribution of each eigenvalue to $B(t)$ is always positive for $L$ even.

\subsubsection{Numerical Checks}
 
The prediction \eqref{eq:traceKI} can be checked by finding numerically all unimodular eigenvalues of $\mathbb{T}$ for short times. The results for times up to $t=25$ are shown in Tab.~\ref{First power}. No eigenvectors are found for odd $t$ while for even $t$ the only eigenvalue is the one given in Eq.~\eqref{eq:eigenvalueT} {(corresponding to the eigenvector \eqref{eq:psi})}. The only exceptions are for $t=6$ and $t=10$. In these two cases we find an additional unit-magnitude eigenvalue
\be
e^{i\varphi}= 1\,,
\label{eq:additionaleigenvalueT}
\ee
and its corresponding eigenvectors have been identified in Ref.~\cite{BKP:SFF} (cf. Eqs.~(171) and (175) of the Supplemental Material). As no other additional eigenvector can be found for ${t>10}$ we conjecture that the presence of \eqref{eq:additionaleigenvalueT} is a short-time fluke.

\begin{table}[t]
		\begin{ruledtabular}
		\begin{tabular}{ c l c l c l c l c l c l c}
			Time  & 2 & 4 & 6 & 8  & 10 & 12 & 14 & 16 & 18 & 20 & 22 & 24 \\ \hline
			$|\lambda|=1$ & -1 & -1 &  $\pm 1$ & -1 & $\pm 1$ & -1 & -1 &-1 & -1 & -1 & -1 & -1\\ 
		\end{tabular}
	\end{ruledtabular}
		\caption{Unit-magnitude eigenvalues $\lambda$ of $\mathbb{T}$ for even times $t\leq24$. There are no such eigenvalues at odd times for $t\leq 25$.}
\label{First power}
\end{table}

\subsection{Higher Moments of the Spectral Form Factor}
\label{Second moment}

Let us now move on to the main objective of this section and consider the moments \eqref{eq:traceT}. The steps to determine a lower bound for these quantities are similar to the ones taken in the previous subsection. In particular, a relation analogue to Eq.~\eqref{eq:property} still holds with $\mathbb T$ and $\mathbb O_{1,0}$ replaced by $\mathbb T_{2 n}$ and $\mathbb O_{n,n}$, namely 
\be
\mathbb T_{2n}^{\dagger}\mathbb T^{\phantom{\dag}}_{2n}=\mathbb{O}^2_{n,n}\,.
\ee
As a consequence, the eigenvalues of $\mathbb T_{2n}$ have again magnitude bounded by one and those with unit magnitude have coinciding algebraic and geometric multiplicity {(while the other eigenvalues remain at a finite distance from the edge of the unit circle, see the right panel of Fig.~\ref{figure_gap} for a representative example)}. Another aspect that is unchanged is that the eigenvectors corresponding to the eigenvalues with unit magnitude belong to the eigenspace of $\mathbb{O}_{n,n}$ with eigenvalue one. This immediately leads to the following two conditions on the relevant (i.e.\ corresponding to unit-magnitude eigenvalues) eigenvectors of $\mathbb T_{2n}$ 
\begin{align}
\Bigl(\mathcal{M}_{z, n}\otimes \1_{{tn}}-\1_{{tn}}\otimes\mathcal{M}_{z, n}\Bigr)|A\rangle &=0\,,
\label{eq:condition 1b}\\
\tilde{\mathcal{U}}_{{\rm KI}, n}^{\phantom{*}}\otimes\tilde{\mathcal{U}}_{{\rm KI}, n}^* |A\rangle &= e^{i\varphi}|A \rangle\,.
\label{eq:condition 2b}
\end{align}
Reasoning along the lines of the previous subsection, one can readily prove that \eqref{eq:condition 1b}--\eqref{eq:condition 2b} are equivalent to 
\begin{align}
\Bigl(\mathcal{M}^{\phantom{*}}_{\alpha, n}\otimes \1_{{tn}}-\1_{{tn}}\otimes\mathcal{M}^*_{\alpha, n}\Bigr)|A\rangle &=0\,,
\label{eq:condition1}\\
\tilde{\mathcal{U}}_{n}^{\phantom{*}}\otimes\tilde{\mathcal{U}}_{n}^* |A\rangle &= e^{i\varphi}|A \rangle\,,
\label{eq:condition2}
\end{align}
where we defined 
\be
\tilde{\mathcal{U}}_{n}\equiv \tilde{U}^{\otimes n}\,.
\ee
Again, using $\tilde{U}^2=\1_{t}$, we have  ${e^{i\varphi}=\pm1}$.

To find a set of eigenvectors  $\{|A\rangle\}$ fulfilling \eqref{eq:condition1}--\eqref{eq:condition2} is useful to follow Ref.~\cite{BKP:SFF} and introduce the a state-to-operator map. This is implemented as follows. First we consider the coefficients $A_{i_1,\ldots,i_{2n}}$ of $|A\rangle$ in the basis 
\be
\{ |i_1, i_2,\cdots, i_{2n-1},i_{2n}\rangle\},
\ee
where $\{|i\rangle\}$ is the computational basis of $\mathcal H_t$. Namely
\be
A_{i_1,\ldots,i_{2n}} \equiv  \braket{i_1, i_2,\cdots, i_{2n-1},i_{2n} |A}
\ee
Then, we define the operator $\mathcal A_n$ in ${\rm End}(\mathcal{H}_t^{\otimes n})$ by means of the following matrix elements 
\be
\braket{i_1\cdots i_n | \mathcal A_n |j_1\cdots j_n} = A_{i_1,\ldots,i_{n},j_1,\ldots,j_n}\,.
\label{eq:operatorstate}
\ee
In this way we can express the conditions \eqref{eq:condition1} and \eqref{eq:condition2} as
\begin{align}
[\mathcal A_n, \mathcal{M}_{\alpha,n}]&=0,
\label{eq:condition 1c}\\
\tilde{\mathcal{U}}_n\mathcal A_n\tilde{\mathcal{U}}^\dag_n &=\pm \mathcal A_n\,.
\label{eq:condition 1d}
\end{align}
The first observation is that, even tough both $+1$ and $-1$ are possible eigenvalues of $\tilde{{U}}$, it is reasonable to restrict ourself to the case of positive eigenvalues. Indeed, as we will see in the following, negative eigenvalues are expected to be rare and appear only for small times. Moreover, considering only positive eigenvalues produces a lower bound for \eqref{eq:traceT} if we only focus on \emph{even lengths}. For this reason, we get rid of the contribution of negative eigenvalues by averaging the results for even and odd lengths, i.e. we define 
\be
\bar {K}_n(t)=\lim_{L\to \infty} \frac{{\rm tr}\left(\mathbb{T}^{2L}_{2n}\right)+{\rm tr}\left(\mathbb{T}^{2L+1}_{2n}\right)}{2}\,. 
\ee
A set of eigenvectors with eigenvalue one can be determined by finding the number of all linearly-independent operators that commute with the set of operators $\{\mathcal{U}_n, \mathcal{M}_{\alpha,n}\}$. This set can be found by observing that, as shown in Ref.~\cite{BKP:SFF}, the elements of the \emph{dihedral group} $\mathcal{G}_t$ commute with the set of operators $\{U, M_{\alpha}\}$. The group $\mathcal{G}_t$ is a symmetry group of a polygon with $t$ vertices and its elements be expressed as 
\be
\{\Pi^p R^m; p\in\{0, t-1\},m\in\{0,1\}\}\,,
\ee
with $\Pi$ denoting the periodic shift for one site and $R$ reflection. These operators are represented in ${\rm End}(\mathcal{H}_t^{\otimes n})$ as 
\be
\Pi=\prod_{\tau=1}^{t-1}P_{\tau, \tau+1} \textnormal{ and } R=\prod_{\tau=1}^{[t/2]}P_{\tau, t+1-\tau}\,,
\ee
where $P_{i,j}$ is the transposition. The number of linearly independent elements of this representation of the dihedral group is~\cite{BKP:SFF}  
\begin{align}
|\mathcal{G}_t|=\begin{cases}
2t,\quad t\geq 6\\
2t-1,\quad t\in\{1,3,4,5\}\\
2,\quad t=2
\end{cases}\,.
\end{align} 
The above facts imply that any operator written as 
\be
\!\mathcal B=\!\!\sum_{m_j=0}^1\,\,\sum_{p_j=0}^{t-1} B_{\boldsymbol p, \boldsymbol m}\,\,\Pi^{p_1}\!R^{m_1}\!\otimes\cdots \otimes \Pi^{p_n}\!R^{m_n}\!,
\label{eq:opB}
\ee
commutes with $\{\mathcal{U}_n, \mathcal{M}_{\alpha,n}\}$. 

This means that the number of operators commuting with $\{\mathcal{U}_n, \mathcal{M}_{\alpha,n}\}$ is at least number of elements of the dihedral group to the power $n$. There is, however, an additional combinatorial prefactor that one should take into account to attain a tighter lower bound. The combinatorial prefactor arises from an arbitrariness in the definition \eqref{eq:operatorstate} of the operator $\mathcal A$. Indeed, it is easy to see that defining 
\be
\braket{i_1\cdots i_n | \mathcal A^{(\tau \sigma)}_n |j_1\cdots j_n} = A_{i_{\tau(1)},j_{\sigma(1)},\ldots,i_{\tau(n)},j_{\sigma(n)}}\,,
\label{eq:operatorstateperm}
\ee
with $\tau,\sigma \in S_{n}$ permutations of $n$ elements, leads to operators fulfilling \eqref{eq:condition 1c}--\eqref{eq:condition 1d} for any $\tau$ and $\sigma$. These operators are not all linearly independent: since the set of all operators $\mathcal B$ (cf. \eqref{eq:opB}) is invariant under permutations of the copies in the tensor product, only $\mathcal A_{1 \sigma}$ can be independent. This leads to a combinatorial prefactor $n!$. Such a combinatorial prefactor leads to a lower bound on the higher moments of the SFF that agrees with the standard COE prediction. 

The fact that $\tilde U=\tilde U^\dag$, however, implies that the combinatorial prefactor is actually higher. Indeed, also 
\be
\braket{i_1\cdots i_n | \bar{\mathcal A}^{(\sigma)}_n |i_{n+1}\cdots i_{2n}} = A_{i_{\sigma(1)},\ldots, i_{\sigma(2n)}}\,,
\label{eq:operatorstateperm2}
\ee
fulfil \eqref{eq:condition 1c}--\eqref{eq:condition 1d} for any permutation of $2n$ elements $\sigma$. To see this we first note that considering the unitary mapping  
\be
\ket{A}\mapsto \ket{A'}=\1_{2^{tn}}\otimes\tilde{\mathcal{F}}_{y, n} \ket{A}
\ee
with 
\be
\tilde{\mathcal{F}}_{y,n}\equiv\underbrace{\tilde{F}_y\otimes\cdots \otimes \tilde{F}_y}_n
\ee
and $\tilde{F}_{y,n}$ defined as in \eqref{eq:FY} with $L$ replaced with $t$, the conditions  \eqref{eq:condition1}--\eqref{eq:condition2} become 
\begin{align}
\Bigl(\mathcal{M}^{\phantom{*}}_{\alpha, n}\otimes \1_{{tn}}+\1_{{tn}}\otimes\mathcal{M}_{\alpha, n}\Bigr)|A'\rangle &=0\,,
\label{eq:condition1mod}\\
\tilde{\mathcal{U}}_{n}^{\phantom{*}}\otimes\tilde{\mathcal{U}}_{n}^* |A'\rangle &= e^{i\varphi}|A' \rangle\,.
\label{eq:condition2mod}
\end{align}
Mapping these into relations for operators by means of the definition \eqref{eq:operatorstateperm2} (with $A$ replaced by $A'$) we then find 
\begin{align}
\{ \bar{\mathcal A'}^{(\sigma)}_n, \mathcal{M}_{\alpha,n}^*\}&=0,
\label{eq:condition 1c}\\
\tilde{\mathcal{U}}_n \bar{\mathcal A'}^{(\sigma)}_n \tilde{\mathcal{U}}^\dag_n&=\pm \bar{\mathcal A'}^{(\sigma)}_n\,.
\label{eq:condition 1d}
\end{align}
Finally, defining 
\be
\bar{\mathcal A}^{(\sigma)}_n= \tilde{\mathcal{F}}_{y,n} \bar{\mathcal A'}^{(\sigma)}_n
\ee
we find that it fulfils \eqref{eq:condition 1c}--\eqref{eq:condition 1d} for all $\sigma\in S_{2n}$. 

Taking again into account the invariance of the set $\{\mathcal B\}$ under permutations of the copies in the tensor product and noting that the set is also invariant under transposition in each single copy we obtain the following combinatorial prefactor 
\be
\frac{(2n)!}{2^n\, n! }=(2n-1)!!\,.
\ee
Together with this additional factor a lower bound for $\bar K_n(t)$ can then be expressed as   
\begin{align}
\bar {K}_n(t) \geq\begin{cases}
	(2t)^n (2n-1)!! , & t\geq 6,\\
	(2t-1)^n (2n-1)!! , & t\in\{1,3,4,5\},\\
	2^n (2n-1)!! , & t=2\,.
\end{cases}
\label{eq:Result}
\end{align}
We see that for {\em odd} times larger than $5$ this bound agrees with the RMT prediction \eqref{eq:cumulantsJacobi} and, therefore, we expect it to be tight. For even times we can find additional operators fulfilling \eqref{eq:condition 1c}--\eqref{eq:condition 1d} by considering $\ket{\psi}\!\!\bra{\psi}$ with $\ket{\psi}$ given in \eqref{eq:psi}. In particular we find the following additional solutions 
\bw
\be
\!\mathcal B^{(k)}=\!\!\sum_{m_j=0}^1\,\,\sum_{p_j=0}^{t-1} B_{\boldsymbol p, \boldsymbol m}\,\,\Pi^{p_1}\!R^{m_1}\!\otimes\cdots \otimes \Pi^{p_k}\!R^{m_k}\otimes\ket{\psi}\!\!\bra{\psi}\cdots\ket{\psi}\!\!\bra{\psi} \!,\qquad\qquad k=0,\ldots,2n-1
\ee
\ew
with a combinatorial prefactor of 
\be
\binom{2n}{2k} (2k-1)!!.
\ee
Taking into account also these solutions we have that the bound agrees with the RMT prediction \eqref{eq:cumulantsJacobi} for all times larger than $6$.

\subsubsection{Numerical Checks}
\label{sec:secondmoment}

The arguments of this section can again be tested (for short times) by identifying numerically all eigenvectors of the space-transfer matrix that have eigenvalues equal to $\pm1$. Here we present an analysis of the simplest non-trivial case, i.e. $n=2$. By repeatedly applying $\mathbb{T}_4$ to a random state and then projecting to different fixed-momentum subspaces ({\em power method}) we enumerated all its unimodular eigenvectors up to $t=7$: the results are gathered in Tab.~\ref{Fourthpower}.  

\begin{table}[t]	
	\begin{ruledtabular}
	\begin{tabular}{ l l l l l l l }
		t                & 2 & 3 & 4 & 5 & 6 & 7 \\ \hline
		Lower bound Eq.~\eqref{eq:Result}& 12 & 75 & 147 & 243 & 432 & 588 \\ \hline
		$\mathcal{N}_{+1}$  & 14 & 59 & 177 & 243 & 507 & 587\\ \hline
		$\mathcal{N}_{-1}$  & 0 & 0 & 4 & 0 & 132 & 0 \\ 
	\end{tabular}
	\end{ruledtabular}
	\caption{Number of eigenvectors of $\mathbb T_4$ with eigenvalue $+1$ or $-1$ obtained via the power method. For comparison, the first row contains the lower bound given in the Eq.~\eqref{eq:Result}.}
	\label{Fourthpower}
\end{table}

The first point to note is that negative eigenvalues are less common than positive ones. For odd times we did not find any eigenvalue $-1$. The next observation is that, as expected, the number of eigenvectors is much bigger than the standard COE prediction. 

However, since we can only investigate the short-time behaviour, we observe some short-time effects that we believe will disappear for larger times. In particular, we observe two main phenomena. First, the number of linearly independent vectors in some subspaces is smaller than expected because vectors are ``not long enough". In other words, for short times the operators identified in the previous section are not all linearly independent. Second, for short even times there are some additional eigenstates (similarly to what happens for $t=6$ and $t=10$ in Sec.~\ref{Half moment}). Since these special states seem to appear only for even times we can avoid this complication by considering only odd times. The first phenomenon, however, remains also there. An example can be readily observed for $t=3$. In this case we find only $59$ eigenvectors with eigenvalue $+1$ even tough the lower bound from Eq.~\eqref{eq:Result} predicts at least $75$ of them. A similar effect can be seen for $t=7$ where we found $587$ eigenvectors, whereas the expected lower bound is higher by one. On the other hand at $t=5$ the number of eigenvectors matches the predicted lower bound. 
 
To obtain more detailed information we note that 
$\mathbb T_4$ commutes with the four translation operators 
\begin{align}
T_1= \Pi \otimes\1\otimes\1\otimes\1,& &T_2= \1\otimes\Pi \otimes\1\otimes\1,\notag\\
T_3=\1\otimes\1\otimes\Pi \otimes\1,& &T_4=\1\otimes\1\otimes\1\otimes\Pi ,
\end{align}
and count how many linearly independent eigenvectors with unit-magnitude eigenvalue exist in each subspace with fixed four-quasi-momentum $\{k_1,k_2,k_3,k_4\}$ (see Appendix~\ref{sec:app} for more details). By analysing the results --- reported in the Tables~\ref{T2}--\ref{T7} --- we identify the following general structure 
\begin{enumerate}
	\setlength\itemsep{0.05 em}
	\item The relevant eigenvectors appear in sectors where four momenta can be arranged into two pairs. Each pair $(k_1,k_2)$ contains two equal momenta $k_1=k_2$, or two momenta in the relation $k_1=t-k_2\equiv-k_2$.
	\item The number of linearly independent vectors in a sector is the same as the number of ways in which four momenta can be grouped into two pairs. This means that we can get the degeneracies one or three in a typical sector. 
	For example 
	\begin{equation}
	\{k_{\tikzmark{rr}1}, k_{\tikzmark{bb}1}, k_{\tikzmark{jj}2}, k_{\tikzmark{faf}2}\},\qquad k_1\neq k_2\neq t-k_2.
	\end{equation}
	\begin{tikzpicture}[overlay,remember picture]
	\draw[latex-latex,shorten <=1pt,shorten >=1pt] 
	($(pic cs:rr)$) 
	-- ++ (0,-.5) -| ($(pic cs:bb)$);
	\draw[latex-latex,shorten <=1pt,shorten >=1pt] 
	($(pic cs:jj)$) 
	-- ++ (0,-.5) -| ($(pic cs:faf)$);
	\draw[latex-latex,shorten <=1pt,shorten >=1pt] 
	($(pic cs:oo)$) 
	-- ++ (0,-.5) -| ($(pic cs:kk)$);
	\end{tikzpicture}
	\be
	\{k_{\tikzmark{kk}1}, k_{\tikzmark{oo}1}, k_{\tikzmark{vv}1}, k_{\tikzmark{pp}1}\},\,\, \{k_{\tikzmark{kkk}1}, k_{\tikzmark{ooo}1}, k_{\tikzmark{vvv}1}, k_{\tikzmark{ppp}1}\},\,\, \{k_{\tikzmark{kkkk}1}, k_{\tikzmark{oooo}1}, k_{\tikzmark{vvvv}1}, k_{\tikzmark{pppp}1}\}.
	\ee
	\begin{tikzpicture}[overlay,remember picture]
	\draw[latex-latex,shorten <=1pt,shorten >=1pt] 
	($(pic cs:oo)$) 
	-- ++ (0,-.5) -| ($(pic cs:kk)$);
	\draw[latex-latex,shorten <=1pt,shorten >=1pt] 
	($(pic cs:pp)$) 
	-- ++ (0,-.6) -| ($(pic cs:vv)$);
	
	\draw[latex-latex,shorten <=1pt,shorten >=1pt] 
	($(pic cs:vvv)$) 
	-- ++ (0,-.5) -| ($(pic cs:kkk)$);
	\draw[latex-latex,shorten <=1pt,shorten >=1pt] 
	($(pic cs:ppp)$) 
	-- ++ (0,-.6) -| ($(pic cs:ooo)$);
	
	\draw[latex-latex,shorten <=1pt,shorten >=1pt] 
	($(pic cs:pppp)$) 
	-- ++ (0,-.5) -| ($(pic cs:kkkk)$);
	\draw[latex-latex,shorten <=1pt,shorten >=1pt] 
	($(pic cs:vvvv)$) 
	-- ++ (0,-.6) -| ($(pic cs:oooo)$);
	\end{tikzpicture}
	
	 The total number of vectors in a sector is therefore always given by the product of two numbers:  the number of all possible pairs and that of all possible permutations of the momenta.  
	\item When a sector has momenta $k/2$ or $0$, one gets independent contributions from even and odd reflection eigenspaces. 
\end{enumerate}
For short times, however, some of the reflection eigenspaces can vanish, or be smaller than expected. For example, at $t=7$ in the reflection odd part of the sector with all four momenta equal to zero, we obtain only $2$ independent vectors instead of the expected three. The same problem occurs for $t=3$ in almost all sectors. The number of sectors where this happens decreased when $t$ increases and this problem is expected to disappear for larger times.

It is interesting to check if by applying the above principles we can calculate the final result for the number of eigenvectors. For (large enough) odd times the result is exactly $12t^2$, while for even times we get $12t^2+12t+1$ (see Appendix~\ref{sec:app}). Both results agree with the lower bound~\eqref{eq:Result} and with the RMT prediction~\eqref{eq:cumulantsJacobi}.

\section{Monte-Carlo simulations}
\label{sec:montecarlo}

In this section we present numerical evidence substantiating the tightness of the bound \eqref{eq:Result}. Our numerical results are obtained by means of simple Monte-Carlo simulations based on direct time propagation with $U_{\rm KI}[\boldsymbol{h}]$ followed by an average over different configurations of the longitudinal magnetic fields $h_j$. 
 
The trace of $U_{\rm KI}^t[\boldsymbol{ h}]$ is computed by restricting the sum to a set $\mathcal{R}$ containing $m$ random states of $\mathbb{C}^{\mathcal N}$. The states ${| \boldsymbol{r}\rangle \in \mathcal{R}}$ are obtained by producing and normalising vectors with independent and identically distributed complex Gaussian random variables. The number of states $m$ can be much smaller than $2^L$ and we expect fluctuations of the order $\mathcal{O}\left({1}/{\sqrt{m}}\right)$.
For example, for $n=2$ the trace is approximated by
\bw
\be\label{tracare}
|tr\left[U^t_{\rm KI}[\boldsymbol{h}]\right]|^4 \approx \frac{2^{4L}}{m(m-1)(m-2)(m-3)} \displaystyle \sum_{\{\boldsymbol{r}_j\}\in\mathcal{R}} \langle \boldsymbol{r}_1 |U^t_{\rm KI}[\boldsymbol{h}]| \boldsymbol{r}_1\rangle \langle \boldsymbol{r}_2 |U^t_{\rm KI}[\boldsymbol{h}]|  \boldsymbol{r}_2\rangle^* \langle \boldsymbol{r}_3 |U^t_{\rm KI}[\boldsymbol{h}]| \boldsymbol{r}_3\rangle \langle \boldsymbol{r}_4 |U_{\rm KI}^t[\boldsymbol{h}]| \boldsymbol{r}_4\rangle^*,
\ee
\ew
and $\boldsymbol{r}_1\neq \boldsymbol{r}_2 \neq \boldsymbol{r}_3\neq \boldsymbol{r}_4$.
The results are obtained for finite-length chains and consequently the thermodynamic limit behaviour can only be observed for times $t< L$.

\begin{figure}[!htb]
	\begin{center}
		\includegraphics[width=.475\textwidth]{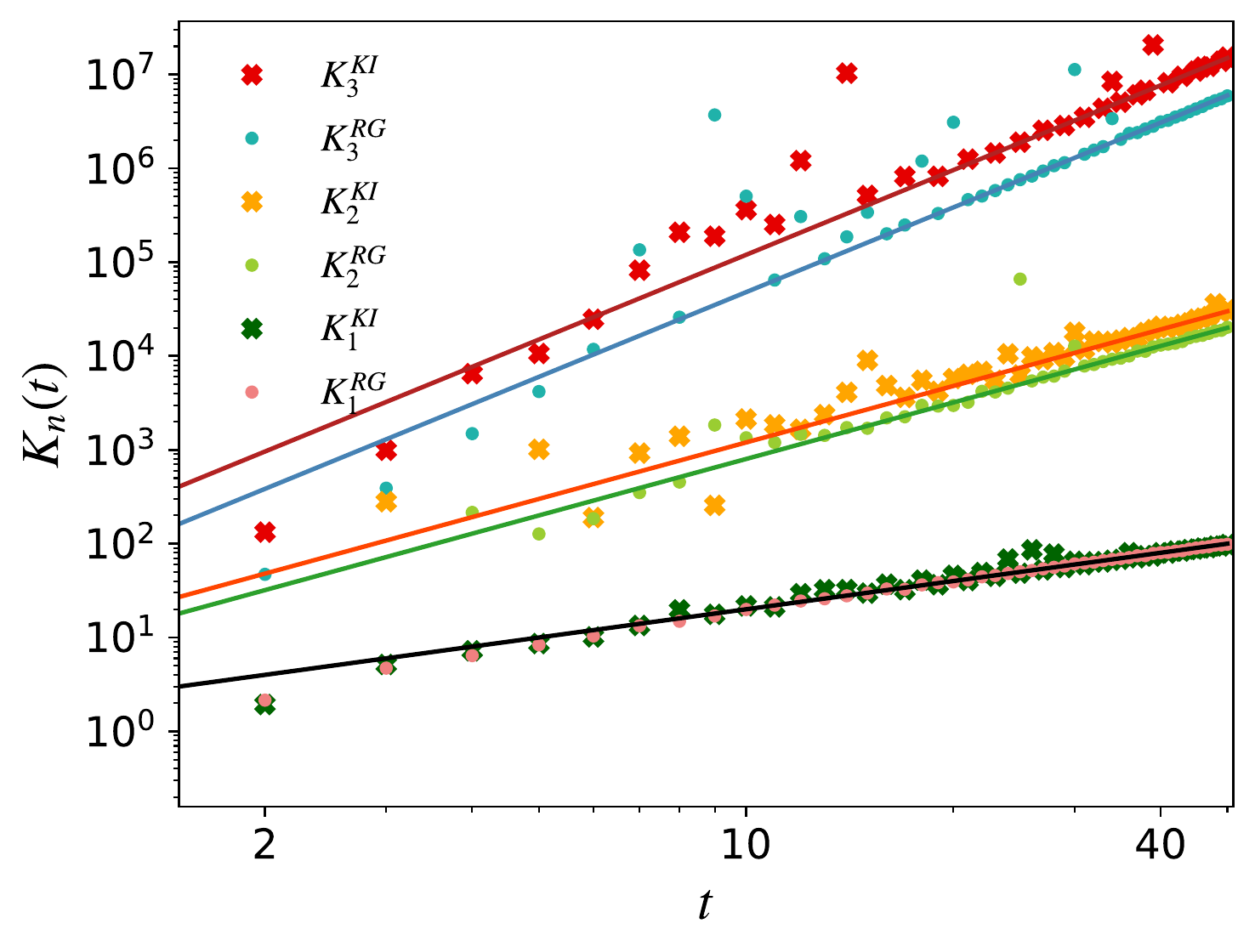}
	\end{center}
	\vspace{-10pt}
	\caption{A comparison between the $K_n(t)$ with $n\in\{1, 2, 3\}$ and the expected results. The solid straight lines are from bottom to top: $y=2t$ (black), $y=8t^2$ (green), $y=12t^2$ (orange), $y=48t^3$ (blue) and $y=120t^3$ (red). The crosses are the data obtained for the self-dual kicked Ising model and the dots represent results for the time-reversal invariant dual-unitary circuits determined by $\phi=J=0$ and $u_{+}=v_-= e^{-ih_j\sigma_z}$, $u_{-}=e^{-i\frac{\pi}{4}\sigma_x}$, $v_{+} =\1_2$. For both models for $K_3(t)$ the system size is $L=13$ and the averaging is done over $516000$ configurations of the fields $\boldsymbol{ h}$ and the trace is computed by definition. For $K_1(t)$ and $K_2(t)$ the system size is $L=15$, $m=128$ and the average is obtained by taking $\approx 200000$ configurations of $\boldsymbol{ h}$. For all $n$ the fields $h_j$ are distributed independently with a Gaussian distribution determined by $\sigma=100\pi$ and $\bar{h}=0.6$.}
	\label{pic2}	
\end{figure}

Fig.~\ref{pic2} reports the results of the Monte-Carlo simulations for ${K}_1(t)$, ${K}_2(t)$ and ${K}_3(t)$. As we see these results indicate that the first, the second and the third moment of the SFF grow with time as predicted by  Eq.~\eqref{eq:Result}. 
{Note that small deviations from the predicted asymptotics are due to finite size effects (we set $L=13$ and $L=15$ in these simulations) which are clearly dominating over the statistical Monte-Carlo errors (of the order
of data point symbol sizes or smaller) and also prohibit to resolve corrections to asymptotics for even times.}

For comparison we also plotted the results for the time-reversal invariant dual-unitary circuits with random gates. The Floquet propagator has the form described in Ref.~\cite{BKP:dual-unitary} (equations (23) and (24)) with $J=0$ and 
\begin{align}
\begin{array}{ll}
u_{+}=v_-= e^{-ih\sigma_z}, \quad u_{-}=e^{-i\frac{\pi}{4}\sigma_x}, \quad v_{+} =\1_2.
\end{array}
\end{align}  
We see that, unlike for the self-dual kicked Ising, the moments agree with the COE predictions.  
 
\begin{figure}[!htb]
	\begin{center}
		
		\includegraphics[width=.475\textwidth]{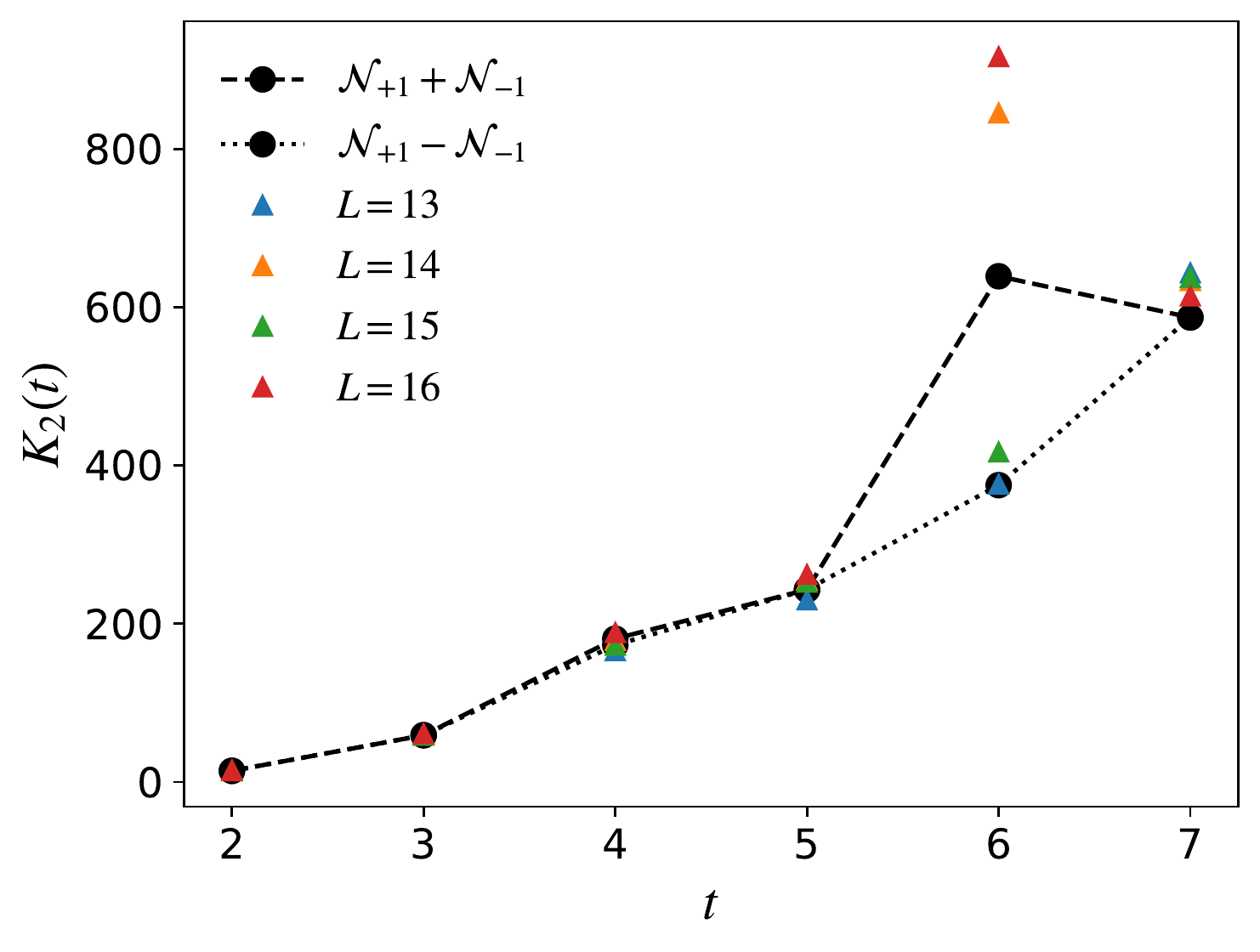}
		
	\end{center}
	\caption{A comparison between the numbers from Tab.~\ref{Fourthpower} (black) and Monte-Carlo simulation (blue, orange, green, red triangles) for $L=13, 14, 15, 16$. The averaging is done over $\approx 20000$ configurations of the magnetic fields $h_j$. The parameters are $\bar{h}=0.6$ and $\sigma=100\pi$. }
	
	\label{pic3}	
\end{figure}

Finally, in order to see whether all eigenvectors are identified, in Figure~\ref{pic3} we compare the Monte-Carlo simulation with the results from Tab.~\ref{Fourthpower}.
The result agrees well for all times except for $t=6$ and $L$ even. This might indicate that some additional eigenvectors with eigenvalues $+1$ and $-1$ are not identified. Other causes of disagreement might be finite-size corrections in the Monte-Carlo simulation or fluctuations due to the finite number of realisations.

\section{Conclusions}
\label{sec:conclusions}

In this paper we computed the statistics of the spectral form factor in the self-dual kicked Ising model. Our strategy has been to establish a rigorous lower bound on the higher moments (generalising the space transfer matrix method of Ref.~\cite{BKP:SFF}) 
and to check its saturation numerically (via Monte-Carlo simulations). We found that, even though the spectral form factor takes the standard COE form, the fluctuations are consistently higher. We explained this result by noting that, since the self dual kicked Ising model has two anti-unitary symmetries~\cite{BWAGG:SFF}, the relevant random matrix ensemble is not the COE but is defined on a more restricted symmetric space. We found that this space is either ${Sp(N)/U(N)}$ or ${O(2N)/{O(N)\!\times\!O(N)}}$ depending on the parity of the number of sites. {Moreover, we found that these ensembles describe the statistics of the spectral form factor in the thermodynamic limit and for all times larger than 6. In particular, this implies that in the self-dual kicked Ising model the Thouless time is $L$-independent and is the same for all cumulants of the spectral form factor.}

Our work suggests several possible directions for future research. An obvious one is to prove rigorously the findings of this paper in the spirit of Ref.~\cite{BKP:SFF}. Namely, devise a mathematical proof of the bound's saturation. Our numerical analysis of the short time behaviour suggests that such a proof is concretely within reach, at least in the case of odd times. 

Moreover, it is interesting to apply the method adopted here to the study of the spectral-form-factor statistics in other systems. Our numerical results, together with recent compelling analytical evidence~\cite{BKP:dual-unitary, BKP:entropy, BKP:OEergodicandmixing, BKP:OEsolitons, GoLa19, PBCP20, CL:OTOCsDU, BP:tripartite, KBP}, suggest that dual-unitary circuits~\cite{BKP:dual-unitary} provide a very convenient framework where these questions can be investigated analytically. Indeed, preliminary results indicate that all circuits in this class are characterised by a vanishing Thouless time, meaning that there is no characteristic time scale other than the Heisenberg time given by the dimension of the Hilbert space. In fact, they seem to provide an arena where one can generate many-different random matrix ensembles by including increasingly more anti-unitary symmetries in the local gates. 

Finally, it is interesting to ask whether the method of this work can be successfully applied to ``generic systems" with non-unitary space transfer matrix. There a meaningful comparison with RMT can only be performed in a finite volume due to a Thouless time increasing monotonically with the volume~\cite{Chalker2, KLP}.   

\section{Acknowledgments}
\label{sec:acknowledgments}
All authors have been supported by the EU Horizon 2020 program through the ERC Advanced Grant OMNES No. 694544, and by the Slovenian Research Agency (ARRS) under the Programme P1-0402. TP acknowledges a fruitful discussion with Nick Hunter-Jones in the preliminary stage of this work.

\appendix

\section{Unimodular eigenvalues of $\mathbb T_4$}
\label{sec:app}

The number of linearly independent eigenvectors associated to unimodular eigenvalues of $\mathbb T_4$ for times $t\in\{2, 3, 4, 5, 6, 7\}$ are reported in the Tables~\ref{T2}--\ref{T7}. Since $\mathbb T_4$ commutes with the four translation operators 
\begin{align}
T_1= \Pi \otimes\1\otimes\1\otimes\1,& &T_2= \1\otimes\Pi \otimes\1\otimes\1,\notag\\
T_3=\1\otimes\1\otimes\Pi \otimes\1,& &T_4=\1\otimes\1\otimes\1\otimes\Pi ,
\end{align}
its eigenvectors can be labelled using four (quasi)momenta $\{k_1,k_2,k_3,k_4\}$. The number of vectors in a sector is the same regardless of the order of the momenta and therefore each combination of four $k$-s is found only once in each table. $P$  (red) is the number of all possible permutations of a certain set of momenta. $D$ (black) is the number of linearly independent vectors in a specific subspace. No additional sign means that only eigenvalues $+1$ are present. If some eigenvalues $-1$ are present, there is a sign $(-)$ beside the number of such eigenvalues and a sign $(+)$ beside the number of eigenvectors belonging to the positive eigenvalue.\par 
By looking at the tables we see a demonstration of the rules described in Sec.~\ref{sec:secondmoment}. To explain results for the special cases where two or four momenta are equal to zero, we note that the states belonging to the reflection symmetric and antisymmetric subspaces are linearly independent for $t\geq 6$. If $0_{+}$ stands for the reflection symmetric subspace and $0_{-}$ the antisymmetric subspace, we expect to find three linearly independent states in the sector $\{0_{-},0_{-},0_{-},  0_{-}\} $, another three in the subspace $\{0_{+}, 0_{+},0_{+}, 0_{+}\}$ and one vector in $\{0_{-},0_{-},0_{+},0_{+}\} $. However, in the last case there are six possible permutations and therefore the total number of linearly independent eigenvectors with $\{0, 0, 0, 0\}$ is twelve. When only two momenta are equal to zero, the number of expected eigenvectors is two. One is in the subspace $\{k, k', 0_{-}, 0_{-}\}$ and the other in $\{k, k', 0_{+}, 0_{+}\}$.\par  
The same happens for even $t$ in sectors with momentum $k=t/2$ because even and odd reflection sectors are both non-trivial for $t\geq 4$. Furthermore, there is the additional state $|\psi\rangle$ (Eq. \eqref{eq:psi})  and it belongs to reflection-symmetric or reflection-antisymmetric subspace depending on parity of $t$. When all four momenta are equal to $t/2$ we expect $25$ linearly independent vectors. By applying the same reasoning as for $k=0$, we get $12$ vectors, the additional $13$ linearly independent vectors contain the state $|\psi\rangle$. When only two momenta are equal to $t/2$, we get three independent vectors. Two of them are due to the same reasons as for $k=0$ and the additional one contains the state $|\psi\rangle$. \par 
All information about different types of sectors and the number of linearly independent vectors is summarised in Tab.~\ref{Tab:sectors}. In the first column we report all possible types of sectors. The column ``{Sectors}" reports the number of sectors of each type and the column ``{Pairings}" contains information about the number of expected linearly independent eigenvectors in the corresponding sector. Finally the column ``{Permutations}" reports the number of possible permutations of the four momenta. In order to obtain the lower bound of the spectral form factor one has to multiply the numbers in each row (choosing $t$ either even or odd) and sum together the results of each row. This method gives the result $12t^2$ for odd times and $12t^2+12t+1$ for even times.

\begingroup 
\squeezetable
\begin{table}
		\begin{ruledtabular}
		\begin{tabular}{ l  l  }
			$(k_1, k_2, k_3, k_4)$ & $D\times P$\\ \hline
			(0, 0, 0, 0) & 3$\quad \times \textbf{1}$ \\  \hline
			(0, 0, 0, 1) & 1$\quad \times \textbf{4}$ \\ \hline
			(0, 0, 1, 1) & 1$\quad \times \textbf{6}$ \\ \hline
			(1, 1, 1, 1) & 1$\quad \times \textbf{1}$ \\ \hline
			(1, 1, 1, 0) & 0 \\ \hline
			TOTAL ($t=2$) & 14 \\ 
		\end{tabular}
	\end{ruledtabular}
	\caption{Eigenvectors with unit eigenvalues for $t=2$.}\label{T2}
\end{table}
\endgroup
\begingroup 
\squeezetable
\begin{table}
		\begin{ruledtabular}
		\begin{tabular}{ l  l  }
			\hline
			$(k_1, k_2, k_3, k_4)$ & $D\times P$\\ \hline
			(0, 0, 1, 2) & 1$\quad \times \textbf{12}$ \\ \hline
			(0, 0, 1, 1) & 1$\quad \times \textbf{6}$ \\ \hline
			(0, 0, 2, 2) & 1$\quad \times \textbf{6}$   \\ \hline
			(1, 1, 1, 1) & 2$\quad \times \textbf{1}$ \\ \hline
			(1, 1, 1, 2) & 2$\quad \times \textbf{4}$   \\ \hline
			(1, 1, 2, 2) & 2$\quad \times \textbf{6}$   \\ \hline
			(2, 2, 2, 1) & 2$\quad \times \textbf{4}$   \\ \hline
			(2, 2, 2, 2) & 2$\quad \times \textbf{1}$   \\ \hline
			(0, 0, 0, 0) & 3 $\quad \times \textbf{1}$  \\ \hline
			TOTAL ($t=3$) & 59 \\ \hline
			
		\end{tabular}
	\end{ruledtabular}
	\caption{Eigenvectors with unit eigenvalues for $t=3$.}
	\label{T3}
\end{table}
\endgroup
\begingroup 
\squeezetable
	\begin{table}
		\begin{ruledtabular}
		\begin{tabular}{ l  l  }
			\hline
			$(k_1, k_2, k_3, k_4)$ & $D\times P$\\ \hline
			(1, 1, 1, 2) & 1$\quad \times \textbf{4}$ \\ \hline
			(3, 3, 3, 2) & 1$\quad \times \textbf{4}$ \\ \hline
			(0, 0, 3, 3) & 1$\quad \times \textbf{6}$ \\ \hline
			(0, 0, 1, 1) & 1$\quad \times \textbf{6}$ \\ \hline
			(1, 2, 3, 3) & 1$\quad \times \textbf{12}$   \\ \hline
			(1, 1, 3, 2) & 1$\quad \times \textbf{12}$ \\ \hline
			(0, 0, 3, 1) & 1$\quad \times \textbf{12}$   \\ \hline
			(0, 0, 2, 2) & 2$\quad \times \textbf{6}$   \\ \hline
			(1, 1, 2, 2) & 2$\quad \times \textbf{6}$   \\ \hline
			(1, 2, 2, 3) & 2$\quad \times \textbf{12}$   \\ \hline
			(2, 2, 3, 3) & 2 $\quad \times \textbf{6}$  \\ \hline
			(1, 3, 1, 1) & 3$\quad \times \textbf{4}$   \\ \hline
			(1, 1, 3, 3) & 3$\quad \times \textbf{6}$  \\ \hline
			(1, 3, 3, 3) & 3$\quad \times \textbf{4}$  \\ \hline
			(1, 1, 1, 1) & 3$\quad \times \textbf{1}$   \\ \hline
			(3, 3, 3, 3) & 3$\quad \times \textbf{1}$ \\ \hline
			(0, 0, 0, 0) & 3$\quad \times \textbf{1}$ \\ \hline
			(2, 2, 2, 2) & 10($+$)+4($-$)$ \times \textbf{1}$  \\ \hline
			TOTAL ($t=4$) & 177($+$) and 4 ($-$) \\ \hline
		\end{tabular}
	\end{ruledtabular}
	\caption{The number of eigenvectors corresponding to the unimodular eigenvalues for $t=4$. In the sector $(2, 2, 2, 2)$, there are four eigenvalues $-1$, which is denoted by ($-$).}\label{T4}
\end{table}
\endgroup
\begingroup 
\squeezetable
\begin{table}
		\begin{ruledtabular}
		\begin{tabular}{ l  l }
			\hline
			($k_1$, $k_2$, $k_3$, $k_4$) & $D\times P$ \\ \hline
			(0, 0, 1, 1) & $1\quad \times \textbf{6}$ \\ \hline 
			(0, 0, 1, 4) & $1\quad \times \textbf{12}$ \\ \hline 
			(0, 0, 2, 2) & $1\quad \times \textbf{6}$ \\ \hline 
			(0, 0, 2, 3) & $1\quad \times \textbf{12}$ \\ \hline 
			(0, 0, 3, 3) & $1\quad \times \textbf{6}$ \\ \hline 
			(0, 0, 4, 4) & $1\quad \times \textbf{6}$ \\ \hline 
			(1, 1, 2, 2) & $1\quad \times \textbf{6}$ \\ \hline 
			(1, 1, 2, 3) & $1\quad \times \textbf{12}$ \\ \hline 
			(1, 1, 3, 3) & $1\quad \times \textbf{6}$ \\ \hline 
			(1, 2, 2, 4) & $1\quad \times \textbf{12}$ \\ \hline 
			(1, 2, 3, 4) & $1\quad \times \textbf{24}$ \\ \hline 
			(1, 3, 3, 4) & $1\quad \times \textbf{12}$ \\ \hline 
			(2, 2, 4, 4) & $1\quad \times \textbf{6}$ \\ \hline 
			(2, 3, 4, 4) & $1\quad \times \textbf{12}$ \\ \hline 
			(3, 3, 4, 4) & $1\quad \times \textbf{6}$ \\ \hline 
			(0, 0, 0, 0) & $3\quad \times \textbf{1}$ \\ \hline 
			(1, 1, 1, 1) & $3\quad \times \textbf{1}$ \\ \hline 
			(1, 1, 1, 4) & $3\quad \times \textbf{4}$ \\ \hline 
			(1, 1, 4, 4) & $3\quad \times \textbf{6}$ \\ \hline 
			(1, 4, 4, 4) & $3\quad \times \textbf{4}$ \\ \hline 
			(2, 2, 2, 2) & $3\quad \times \textbf{1}$ \\ \hline 
			(2, 2, 2, 3) & $3\quad \times \textbf{4}$ \\ \hline 
			(2, 2, 3, 3) & $3\quad \times \textbf{6}$ \\ \hline 
			(2, 3, 3, 3) & $3\quad \times \textbf{4}$ \\ \hline 
			(3, 3, 3, 3) & $3\quad \times \textbf{1}$ \\ \hline 
			(4, 4, 4, 4) & $3\quad \times \textbf{1}$ \\ \hline
			TOTAL ($t=5$)  & 243 \\ \hline
		\end{tabular}
	\end{ruledtabular}
	\label{tab5}
	\caption{The number of eigenvectors corresponding to the unimodular eigenvalues for $t=5$.}
\end{table}
\endgroup

\begingroup
\squeezetable
\begin{table}
	\begin{ruledtabular}
	\begin{tabular}{ l  l || l  l }
		\hline
		($k_1$, $k_2$, $k_3$, $k_4$) & $D\times P$ & ($k_1$, $k_2$, $k_3$, $k_4$) & $D\times P$ \\ \hline
		(1, 1, 2, 2) & $1\quad \times \textbf{6}$ & (2, 2, 4, 4) & $3\quad \times \textbf{6}$ \\ \hline 
		(1, 1, 2, 4) & $1\quad \times \textbf{12}$ & (2, 3, 3, 4) & $3\quad \times \textbf{12}$ \\ \hline 
		(1, 1, 4, 4) & $1\quad \times \textbf{6}$ & (2, 4, 4, 4) & $3\quad \times \textbf{4}$  \\ \hline 
		(1, 2, 2, 5) & $1\quad \times \textbf{12}$ & (3, 3, 4, 4) & $3\quad \times \textbf{6}$ \\ \hline 
		(1, 2, 4, 5) & $1\quad \times \textbf{24}$ & (3, 3, 5, 5) & $3\quad \times \textbf{6}$ \\ \hline 
		(1, 4, 4, 5) & $1\quad \times \textbf{12}$ &(4, 4, 4, 4) & $3\quad \times \textbf{1}$ \\ \hline 
		(2, 2, 5, 5) & $1\quad \times \textbf{6}$ & (5, 5, 5, 5) & $3\quad \times \textbf{1}$   \\ \hline 
		(2, 4, 5, 5) & $1\quad \times \textbf{12}$ & (0, 0, 0, 0) & 10 $\quad \times \textbf{1}$    \\ \hline 
		(4, 4, 5, 5) & $1\quad \times \textbf{6}$ & (0, 0, 3, 3) & 6 $\quad \times \textbf{6} $   \\ \hline  
		(0, 0, 1, 5) & $2\quad \times \textbf{12}$ & (3, 3, 3, 3) & $25(+) + 4(-) \times \textbf{1} $    \\ \hline 
		(0, 0, 2, 2) & $2\quad \times \textbf{6}$ & (0, 3, 3, 3) & 4(-) +1(+) $\quad \times \textbf{4} $   \\ \hline 
		(0, 0, 2, 4) & $2\quad \times \textbf{12}$ & (0, 0, 0, 3) & 4(-) $\quad \times \textbf{4} $   \\ \hline 
		(0, 0, 4, 4) & $2\quad \times \textbf{6}$ & (2, 2, 3, 3) & $3\quad \times \textbf{6}$    \\ \hline 
		(0, 0, 5, 5) & $2\quad \times \textbf{6}$ &  (2, 2, 2, 4) & $3\quad \times \textbf{4}$    \\ \hline 
		(0, 0, 1, 1) & $2\quad \times \textbf{6}$ & (2, 2, 2, 2) & $3\quad \times \textbf{1}$     \\ \hline   
		(1, 1, 1, 1) & $3\quad \times \textbf{1}$ &  (1, 5, 5, 5) & $3\quad \times \textbf{4}$    \\ \hline 
		(1, 1, 1, 5) & $3\quad \times \textbf{4}$ & (1, 3, 3, 5) & $3\quad \times \textbf{12}$    \\ \hline 
		(1, 1, 3, 3) & $3\quad \times \textbf{6}$ & (1, 1, 5, 5) & $3\quad \times \textbf{6}$     \\ \hline 
		(0, 3, 1, 1) & $1(-)\quad \times \textbf{12}$ & (0, 3, 2, 2) & $1(-)\quad \times \textbf{12}$ \\ \hline
		(0, 3, 4, 4) & $1(-)\quad \times \textbf{12}$ & (0, 3, 5, 5) & $1(-)\quad \times \textbf{12}$ \\ \hline 
		(0, 3, 1, 5) & $1(-)\quad \times \textbf{24}$ & (0, 3, 2, 4) & $1(-)\quad \times \textbf{24}$ \\ \hline
		&  &   TOTAL ($t=6$) & 507(+) + 132(-)\\ \hline  
	\end{tabular}
\end{ruledtabular}
	\caption{The number of eigenvectors corresponding to the unimodular eigenvalues for $t=6$.}\label{T6}
\end{table}
\endgroup

\begingroup
\squeezetable
\begin{table}[H]
	\begin{ruledtabular}
	\begin{tabular}{ l  l || l  l }
		\hline
		($k_1$, $k_2$, $k_3$, $k_4$) & $D\times P$ & ($k_1$, $k_2$, $k_3$, $k_4$) & $D\times P$ \\ \hline
		(1, 1, 2, 2) & $1\quad \times \textbf{6}$ & (0, 0, 0, 0) & $11 \quad \times \textbf{1}$ \\ \hline 
		(1, 1, 2, 5) & $1\quad \times \textbf{12}$ & (0, 0, 1, 1) & $2\quad \times \textbf{6}$\\ \hline 
		(1, 1, 3, 3) & $1\quad \times \textbf{6}$ & (0, 0, 2, 2) & $2\quad \times \textbf{6}$  \\ \hline 
		(1, 1, 3, 4) & $1\quad \times \textbf{12}$ & (0, 0, 3, 3) & $2\quad \times \textbf{6}$ \\ \hline 
		(1, 1, 4, 4) & $1\quad \times \textbf{6}$ & (0, 0, 4, 4) & $2\quad \times \textbf{6}$ \\ \hline 
		(1, 1, 5, 5) & $1\quad \times \textbf{6}$ & (0, 0, 5, 5) & $2\quad \times \textbf{6}$  \\ \hline 
		(1, 2, 2, 6) & $1\quad \times \textbf{12}$ &  (0, 0, 6, 6) & $2\quad \times \textbf{6}$    \\ \hline 
		(1, 2, 5, 6) & $1\quad \times \textbf{24}$ &  (6, 6, 6, 6) & $3\quad \times \textbf{1}$    \\ \hline 
		(1, 3, 3, 6) & $1\quad \times \textbf{12}$ &  (5, 5, 5, 5) & $3\quad \times \textbf{1}$   \\ \hline 
		(1, 3, 4, 6) & $1\quad \times \textbf{24}$ &  (4, 4, 4, 4) & $3\quad \times \textbf{1}$    \\ \hline 
		(1, 4, 4, 6) & $1\quad \times \textbf{12}$ &  (3, 4, 4, 4) & $3\quad \times \textbf{4}$    \\ \hline 
		(1, 5, 5, 6) & $1\quad \times \textbf{12}$ &  (3, 3, 4, 4) & $3\quad \times \textbf{6}$    \\ \hline 
		(2, 2, 3, 3) & $1\quad \times \textbf{6}$ &  (3, 3, 3, 4) & $3\quad \times \textbf{4}$    \\ \hline 
		(2, 2, 3, 4) & $1\quad \times \textbf{12}$ &  (3, 3, 3, 3) & $3\quad \times \textbf{1}$    \\ \hline 
		(2, 2, 4, 4) & $1\quad \times \textbf{6}$ &  (2, 5, 5, 5) & $3\quad \times \textbf{4}$    \\ \hline 
		(2, 2, 6, 6) & $1\quad \times \textbf{6}$ &  (2, 2, 5, 5) & $3\quad \times \textbf{6}$    \\ \hline 
		(2, 3, 3, 5) & $1\quad \times \textbf{12}$ &  (2, 2, 2, 5) & $3\quad \times \textbf{4}$     \\ \hline 
		(2, 3, 4, 5) & $1\quad \times \textbf{24}$ &  (2, 2, 2, 2) & $3\quad \times \textbf{1}$    \\ \hline 
		(2, 4, 4, 5) & $1\quad \times \textbf{12}$ &  (1, 6, 6, 6) & $3\quad \times \textbf{4}$    \\ \hline 
		(2, 5, 6, 6) & $1\quad \times \textbf{12}$ &  (1, 1, 6, 6) & $3\quad \times \textbf{6}$    \\ \hline 
		(3, 3, 5, 5) & $1\quad \times \textbf{6}$ &  (1, 1, 1, 6) & $3\quad \times \textbf{4}$    \\ \hline 
		(3, 3, 6, 6) & $1\quad \times \textbf{6}$ &  (1, 1, 1, 1) & $3\quad \times \textbf{1}$    \\ \hline 
		(3, 4, 5, 5) & $1\quad \times \textbf{12}$ &  (0, 0, 3, 4) & $2\quad \times \textbf{12}$    \\ \hline 
		(3, 4, 6, 6) & $1\quad \times \textbf{12}$ &  (0, 0, 2, 5) & $2\quad \times \textbf{12}$    \\ \hline 
		(4, 4, 5, 5) & $1\quad \times \textbf{6}$ &  (0, 0, 1, 6) & $2\quad \times \textbf{12}$    \\ \hline 
		(4, 4, 6, 6) & $1\quad \times \textbf{6}$ &  (5, 5, 6, 6) & $1\quad \times \textbf{6}$    \\ \hline 
		& & TOTAL ($t=7$) & 587 \\ \hline
	\end{tabular}
\end{ruledtabular}
\caption{Eigenvectors belonging to the unit eigenvalues for $t =7$. } 
	\label{T7}
\end{table}
\endgroup

\begin{table*}[t]
	\begin{ruledtabular}
		\begin{tabular}{l l| l l| l l| l}
			\multicolumn{2}{l|}{Type}                & \multicolumn{2}{l|}{Sectors} & \multicolumn{2}{l|}{Pairings} & Permutations        \\ \hline
			\multicolumn{2}{l|}{$k,k'\notin\{0,t/2\}$, $k\neq k'$ and $k\neq -k'$}                     & $t$ odd              & $t$ even             &     $t$ odd            &       $t$ even   &              \\ \hline
			\multirow{3}{*}{$\{\bullet,\bullet,\bullet,\bullet\}$}    & $\{0,0,0,0\}$           & $1$                & $1$               &       $12$       & $12$ & \multirow{3}{*}{$1$}  \\ 
			& $\{t/2, t/2, t/2, t/2\}$ & /                & $1$               &    /            &       $27$          &    \\ 
			& $\{k,k,k,k\}$         & $t-1$              & $t-2$             &       $3$    &  $3$   &                    \\ \hline 
			$\{\bullet,\bullet,\bullet,\bullet'\}$                    & $\{k,k,k,-k\}$         & $t-1$              & $t-2$             &     $3$     &  $3$    & 4                   \\ \hline
			\multirow{5}{*}{$\{\bullet,\bullet,\bullet',\bullet'\}$}  & $\{k,k,-k,-k\}$         & $\frac{t-1}{2}$            & $\frac{t-2}{2}$           &      $3$       & $3$  & \multirow{5}{*}{6}  \\ 
			& $\{k,k,0,0\}$           &     $t-1$             &       $t-2$          &          $2$      &        $2$       &      \\ 
			& $\{t/2,t/2,k,k\}$       &          /        &         $t-2$        &       /         &     $3$          &      \\ 
			& $\{t/2,t/2,0,0\}$       &         /         &        $1$         &        /        &        $6$       &      \\ 
			& $\{k,k,k',k'\}$         &        $\frac{(t-1)(t-3)}{2}$          &         $\frac{(t-2)(t-4)}{2}$        &      $1$          &      $1$         &      \\ \hline
			\multirow{3}{*}{$\{\bullet,\bullet,\bullet',\bullet''\}$} & $\{0,0,k,-k\}$          &        $\frac{t-1}{2}$          &        $\frac{t-2}{2}$         &      $2$        & $2$ & \multirow{3}{*}{12} \\ 
			& $\{t/2,t/2,k,-k\}$      &         /         &        $\frac{t-2}{2}$         &     /           &       $3$         &     \\ 
			& $\{k,k,k',-k'\}$         &        $\frac{(t-1)(t-3)}{2}$          &       $\frac{(t-2)(t-4)}{2}$          &      $1$          &     $1$           &     \\ \hline
			$\{\bullet,\bullet',\bullet'',\bullet'''\}$                & $\{k,-k,k',-k'\}$         &         $\frac{(t-1)(t-3)}{8}$         &       $\frac{(t-2)(t-4)}{8}$          &    $1$     &  $1$     & $24$             \\ \hline
		\end{tabular}
	\end{ruledtabular}
	\caption{The table contains the information about different types of sectors. The column ``Sectors" reports the number of sectors with non-zero eigenvectors associated to unimodular eigenvalues are of a certain type. The column ``{Pairings}", reports the number of ways in which one can construct two pairs in a sector. The column ``Permutations" contains the number possible permutations.}\label{Tab:sectors}
\end{table*}

\end{document}